\documentclass[sigconf]{acmart}

\AtBeginDocument{%
  }

\setcopyright{acmlicensed}
\copyrightyear{2018}
\acmYear{2018}
\acmDOI{XXXXXXX.XXXXXXX}

\acmISBN{978-1-4503-XXXX-X/2018/06}
\usepackage{makecell}

\usepackage{bm}
\usepackage{amsmath}
\usepackage{enumitem}
\usepackage{etoolbox}
\usepackage{multirow}
\usepackage{color,colortbl}
\usepackage{tikz}
\usepackage{pgfplots}
\pgfplotsset{compat=1.18}
\usepackage{float}

\usepackage{titlesec}

\usepackage{xcolor}
\definecolor{mylightblue}{rgb}{0.5, 0.7, 1}

\usepackage{pgfplots}
\usepgfplotslibrary{groupplots}

\usepackage[most]{tcolorbox}
\newtcolorbox{AIbox}[1]{colback=gray!5!white,colframe=black!75!black,title=#1}

\appto{\ttfamily}{\hyphenchar\font=`\-}
\usepackage[noend]{algpseudocode}
\algrenewcommand\algorithmicforall{\textbf{for each}}
\newcommand{\ForEach}[2]{\ForAll{#1} #2}
\usepackage[ruled,linesnumbered,vlined]{algorithm2e}

\newtheorem{problem}{Problem}

\titlespacing*{\subsection}{0pt}{2.5pt plus 0.5pt minus 0.5pt}{2.5pt plus 0.5pt minus 0.5pt}

\begin{document}

\title{Clue-RAG: Towards Accurate and Cost-Efficient Graph-based RAG via Multi-Partite Graph and Query-Driven Iterative Retrieval}

\author{Yaodong Su}
\email{yaodongsu@link.cuhk.edu.cn}
\affiliation{%
  \institution{The Chinese University of Hong Kong, Shenzhen}
 \country{}
}

\author{Yixiang Fang}
\authornote{Yixiang Fang is the corresponding author.}
\email{fangyixiang@cuhk.edu.cn}
\affiliation{
  \institution{The Chinese University of Hong Kong, Shenzhen}
  \country{}
}

\author{Yingli Zhou}
\email{yinglizhou@link.cuhk.edu.cn}
\affiliation{
  \institution{The Chinese University of Hong Kong, Shenzhen}
  \country{}
}

\author{Quanqing Xu}
\email{xuquanqing.xqq@oceanbase.com}
\affiliation{
  \institution{OceanBase, Ant Group}
  \country{}
}

\author{Chuanhui Yang}
\email{rizhao.ych@oceanbase.com}
\affiliation{
  \institution{OceanBase, Ant Group}
  \country{}
}

\renewcommand{\shortauthors}{Yaodong Su et al.}

\begin{abstract}
Despite the remarkable progress of Large Language Models (LLMs), their performance in question answering (QA) remains limited by the lack of domain-specific and up-to-date knowledge. Retrieval-Augmented Generation (RAG) addresses this limitation by incorporating external information, often from graph-structured data. However, existing graph-based RAG methods suffer from poor graph quality due to incomplete extraction and insufficient utilization of query information during retrieval. To overcome these limitations, we propose Clue-RAG, a novel approach that introduces (1) a multi-partite graph index incorporates 
\textit{text \underline{C}hunk},
\textit{know\underline{l}edge \underline{u}nit}, and
\textit{\underline{e}ntity}
to capture semantic content at multiple levels of granularity, coupled with a hybrid extraction strategy that reduces LLM token usage while still producing accurate and disambiguated knowledge units, and (2) {\ttfamily Q-Iter}, a query-driven iterative retrieval strategy that enhances relevance through semantic search and constrained graph traversal. Experiments on three QA benchmarks show that Clue-RAG significantly outperforms state-of-the-art baselines, achieving up to 99.33\% higher Accuracy and 113.51\% higher F1 score while reducing indexing costs by 72.58\%. Remarkably, Clue-RAG matches or outperforms baselines even without using an LLM for indexing. These results demonstrate the effectiveness and cost-efficiency of Clue-RAG in advancing graph-based RAG systems. The source code is available in \textcolor{mylightblue}{\url{https://github.com/Feesuu/ClueRAG}}.

\end{abstract}


\maketitle

\section{Introduction}
\label{introduction}

Large Language Models (LLMs) like Qwen3 \cite{yang2025qwen3}, DeepSeek \cite{deepseekai2025deepseekv3technicalreport}, and LLaMA3.1 \cite{grattafiori2024llama} have received tremendous attention from both industry and academia \cite{ghimire2024generative,huang2024survey,liu2024survey,nie2024survey,wang2024large}. Despite their remarkable success in question-answering (QA) tasks, they may still generate wrong answers due to a lack of domain-specific and real-time updated knowledge outside their pre-training corpus \cite{peng2024graph}. To enhance the trustworthiness and interpretability of LLMs, Retrieval-Augmented Generation (RAG) methods \cite{fan2024survey,gao2023retrieval,hu2024rag,huang2024survey,wu2024retrieval,yu2024evaluation,zhao2024retrieval, wu2024medical} have emerged as a core approach, which often retrieve relevant information from documents, relational data, and graph data to facilitate the QA tasks.
The state-of-the-art (SOTA) approaches often use the graph to model the external data since they capture the rich semantic information and link relationships between entities. 

\begin{figure}[t]
    \centering
    \includegraphics[width=\linewidth]{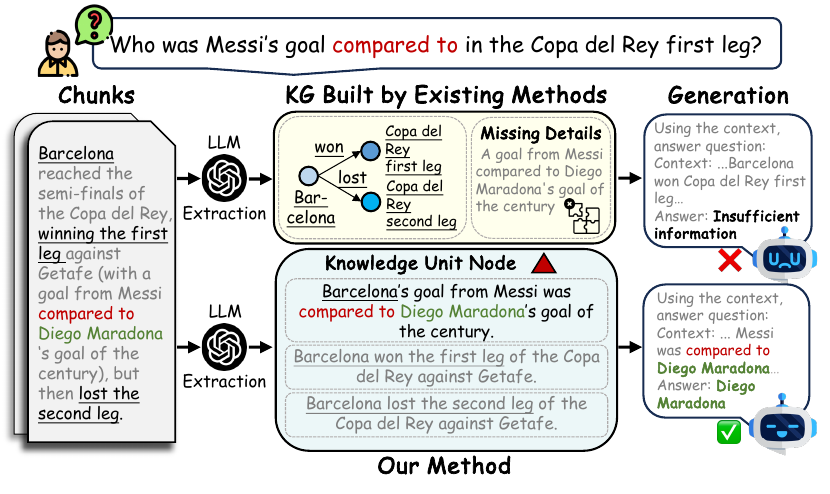}
    \caption{Comparison of triple extraction in existing Graph-based RAG methods (Upper) and ours (Below).}
    \label{fig:small_demo}
\end{figure}

In the literature, representative graph-based RAG methods typically follow a two-phase pipeline:
(1) {\it Offline index construction}: They first segment the external text corpora \(\mathcal{T}\) into small chunks, then extract the nodes and edges, together with their associated textual attributes (e.g., descriptions \cite{edge2024local} or keywords \cite{guo2024lightrag}) from chunks using LLMs, and finally build a knowledge graph (KG) \cite{zhou2025depth, xiang2025use}.
(2) {\it Online retrieval}: Given an online query $q$, they either retrieve relevant nodes, edges, or subgraphs from the KG, or optionally trace these elements back to their source text chunks via associated provenance links \cite{gutierrez2024hipporag, edge2024local, guo2024lightrag}.
Afterwards, they incorporate the relevant information from KG or chunks into a prompt template, and then feed the prompt into an LLM for answer generation.
Although these methods offer improved performance, they still suffer from two major limitations.

$\bullet$ \textbf{Limitation 1: the indexes suffer from the incompleteness issue.}
As shown in recent studies \cite{zhou2025depth,xiang2025use}, the constructed KGs often suffer from incompleteness, typically due to missing key nodes, edges, or the omission of information that cannot be incorporated into structured graph elements.
Take the text chunk in Figure~\ref{fig:small_demo} as an example.
Existing methods typically extract only a few triples (e.g., $\langle \textit{Barcelona}, \textit{won}, \textit{Copa del Rey}$ $\textit{ first legagainst Getafe} \rangle$), while completely ignoring valuable textual information, such as nuanced comparisons between Messi's performance and Maradona's ``Goal of the Century''.
This incompleteness may lead to retrieval errors, i.e., when a query involving this detail is issued, the KG fails to provide sufficient information for generating an accurate answer, as shown in Figure~\ref{fig:small_demo}.

$\bullet$ \textbf{Limitation 2: the retrieval methods face the issue of semantic misalignment.}
In the online retrieval phase, most existing graph-based RAG methods \cite{huang2025ket, edge2024local, guo2024lightrag, gutierrez2024hipporag} follow two steps:
(1) use the input query to identify relevant elements (e.g., nodes, edges) in the KG, and (2) retrieve additional information based on the above elements, such as text chunks or subgraphs, to augment the context for answer generation.
While step (1) ensures query relevance, step (2) often fails to fully leverage the query information, leading to a semantic misalignment between the query and the retrieved context.

\begin{figure*}[htbp]
    \centering
    \includegraphics[width=\textwidth]{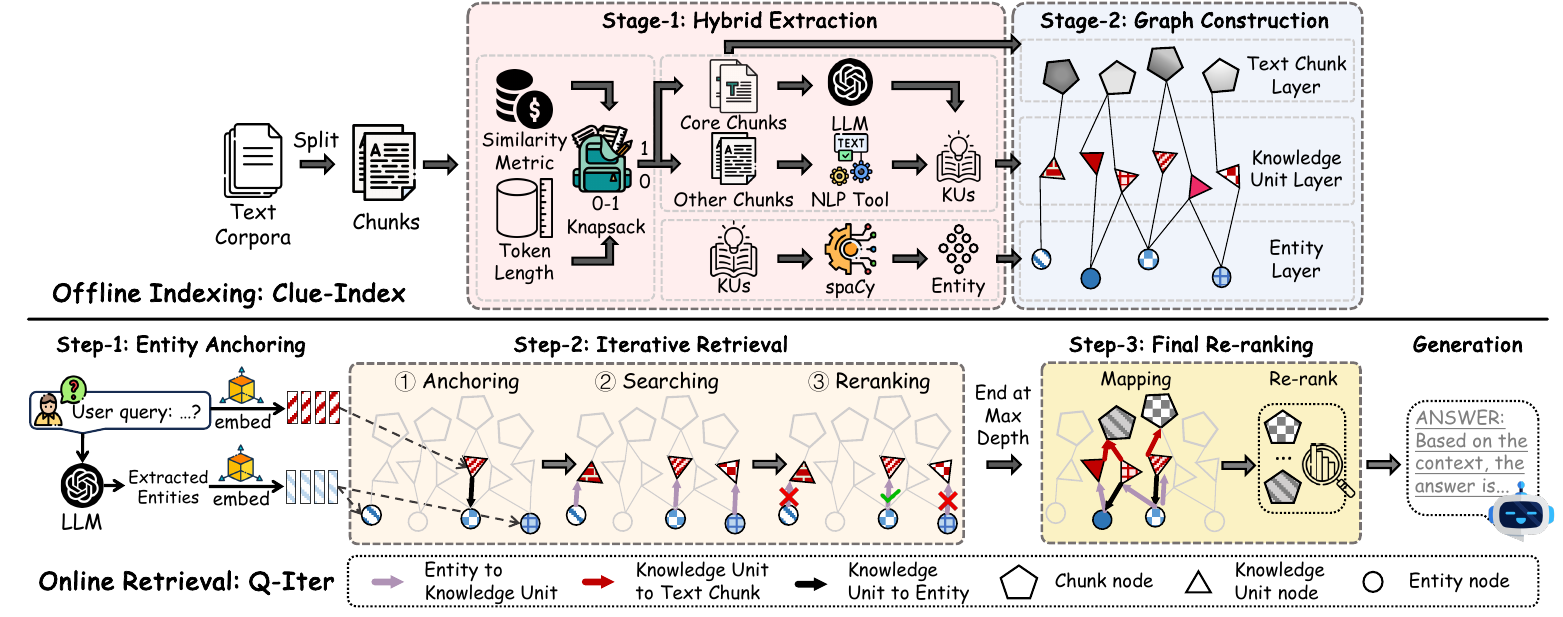}
    \caption{Clue-RAG consists of two phases: offline indexing and online retrieval.}
    \label{fig:overview}
\end{figure*}

To address the above limitations, this paper introduces Clue-RAG, a novel graph-based RAG approach, which not only builds a high quality graph, but also well utilize the query information for answering the questions.
For \textbf{Limitation 1}, we introduce a novel concept, called \textit{knowledge unit}, which is a statement that conveys an atomic piece of information from a text chunk.
Collectively, these knowledge units can fully reconstruct the semantic content of the original text.
By extracting entities from the knowledge units, we can effectively address missing nodes or edges in the constructed KG.
As a result, we build a multi-partite graph composed of three types of nodes, each representing distinct semantic granularities:
\textit{text \underline{C}hunk},
\textit{know\underline{l}edge \underline{u}nit}, and
\textit{\underline{e}ntity}.
These nodes are connected through extraction and containment relations, forming a coherent multi-partite graph, as illustrated in Section \ref{Clue-index} and Figure \ref{fig:overview}.

\begin{figure}[htb]
    \centering
    \includegraphics[width=0.81\linewidth]{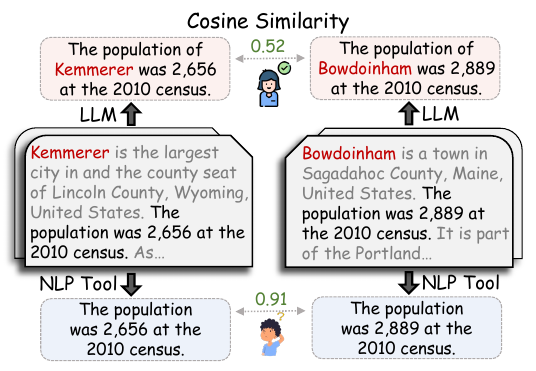}
    \caption{Comparing the knowledge units extracted by NLP tool and LLM.}
    \label{fig:demo}
\end{figure}

Intuitively, we can use some lightweight NLP tools~\cite{Bird_Natural_Language_Processing_2009} to extract sentences from a text chunk, and each sentence can serve as a knowledge unit.
However, this approach overlooks sentence context, leading to contextual ambiguity -- semantically similar sentences from different chunks may actually have distinct meanings.
In Figure~\ref{fig:demo}, for example, if we use an NLP tool to extract sentences from the two chunks, the results may have high cosine similarity (0.91) despite differing in meaning.
To address this, we leverage LLMs to incorporate context when extracting knowledge units. As a result, the LLM-extracted units in the same example show much lower similarity (0.51), effectively capturing context-sensitive distinctions.
However, directly using LLMs to extract the units for all text chunks is very time and token-consuming. 

In this paper, we propose a novel hybrid extraction strategy that combines the advantages of powerful LLMs and lightweight NLP tools \cite{Bird_Natural_Language_Processing_2009}.
An ideal strategy would involve an oracle model that can identify which knowledge unit in a chunk may exhibit contextual ambiguity with units in other chunks, and then use an LLM to resolve the ambiguous unit, while relying on NLP tools for unambiguous units.
However, such an oracle model does not exist in practice.
In light of this, we propose to use semantic similarity between chunks as a proxy to evaluate potential contextual ambiguity, since knowledge units in highly similar chunks are intuitively more prone to ambiguity.
Specifically, we assign chunks having high semantic similarity with others for LLMs processing, while the remainder use NLP tools.
To balance token cost and extraction accuracy, we formulate the chunk assignment as a 0-1 knapsack problem \cite{sahni1975approximate}, aiming to maximize the total semantic similarity of the chunks selected for LLM processing within a given token budget.
As shown in our later experiments, this strategy matches the performance of end-to-end LLM-based unit extraction, but only uses 50\% of the tokens.
Remarkably, Clue-RAG matches or outperforms baselines even without using an LLM for indexing.

To address \textbf{Limitation 2}, we propose \underline{Q}uery-driven \underline{ite}rative \underline{r}etrieval ({\ttfamily Q-Iter}), an iterative retrieval strategy that alternately traverses knowledge units and entities. The process begins by extracting entities from the query to activate related entities, while concurrently using semantic search to anchor associated knowledge units. These units, in turn, expand the entity set by introducing entities referenced in their content. The strategy then performs constrained multihop traversal over the multi-partite graph to retrieve more relevant units. To ensure alignment with the query, each candidate unit is scored using a lightweight re-ranker specialized in query-context relevance. Finally, the collected knowledge units are mapped to their originating text chunks, which are subsequently re-ranked to yield the final context for answer generation.

In our experiments, we test 13 solutions across three datasets, focusing on two key aspects:
(1) \textit{Cost efficiency}, based on token usage during offline indexing and online retrieval; and (2) \textit{QA performance}, measured by F1 score and Accuracy.
Remarkably, Clue-RAG demonstrates significant improvements over the SOTA baselines, delivering up to 99.33\% higher Accuracy and 113.51\% greater F1 score while simultaneously reducing indexing costs by 72.58\%.
Besides, the zero-token variant of Clue-RAG achieves comparable or superior performance to baselines.
This highlights Clue-RAG's superior graph indexing and inherent retrieval capabilities.

In summary, we make the following principal contributions:
\begin{itemize}
    

   \item We propose a novel multi-partite graph index that integrates text chunks, knowledge units, and entities to enable semantically coherent cross-granularity retrieval, coupled with a hybrid extraction strategy for knowledge units that strategically balances the benefits of LLM processing against token usage.
    
    \item We propose {\ttfamily Q-Iter}, an iterative retrieval strategy that incrementally expands relevant knowledge units through semantic search and constrained graph traversal, significantly improving RAG performance.

    \item We conduct extensive experiments and demonstrate the superior performance of Clue-RAG over baselines.
\end{itemize}


{\bf Outline.} We provide the preliminaries and problem formulation in Section \ref{sec:preliminary}.
Section \ref{overview} presents an overview of Clue-RAG, Section \ref{Clue-index} details the construction of the offline index {\ttfamily Clue-Index}, and Section \ref{Q-Iter} describes the online retrieval algorithm {\ttfamily Q-Iter}.
The experimental results are then reported in Section \ref{sec:experiments}.
We review related work in Section \ref{sec:related} and conclude the paper in Section \ref{sec:conclude}.

\section{Preliminary}
\label{sec:preliminary}

Let \(\mathcal{T}\) be a text corpora whose element \(t_i \in \mathcal{T}\) is a text chunk whose token length is at most $L$.
The embedding of each text chunk \(t_i\) is denoted by \(\bm{\phi}(t_i)\).
We denote a graph by \(\mathcal{G}=(\mathcal{V,E})\), where \(\mathcal{V} \) represents the set of nodes, and \(\mathcal{E}\subseteq \mathcal{V} \times \mathcal{V}  \) denotes the set of edges connecting pairs of nodes.
In addition, we use calligraphic uppercase letters (e.g., \(\mathcal{S}\)) to represent a set of elements.
Table \ref{tab:notation} provides a summary of frequently used notations.

\begin{table}[htbp]
  \caption{Notations and meanings.}
  \label{tab:notation}
  \centering
  \small
  \begin{tabular}{ccl}  
    \hline
    \textbf{Category} & \textbf{Notation} & \textbf{Meaning} \\
    \hline
    \multirow{4}{*}{\makecell{Text \\ Processing}} 
    & \(\mathcal{R}(\cdot,\cdot)\) & The re-ranking function \\
    & $\bm{\phi}(t_i)$ & The embedding of text $t_i$ \\ 
    & \(\oplus(\mathcal{S})\) & Concatenation of texts in set \(\mathcal{S}\) \\
    & $\mathcal{T}, t_i, L$ & Text $t_i \in \mathcal{T}$ whose length is $\leq L$ \\ 
    \hline
    \multirow{5}{*}{Retrieval} 
    & $M$ & Beam size $M$ \\
    & $D$ & Search depth $D$ \\
    & $N$ & Returned chunks size $N$ \\
    & $K$ & Top-$K$ relevant results $K$ \\
    & $\alpha$ & Token constraint coefficient $\alpha$ \\
    \hline
    \multirow{1}{*}{Index} 
    & \makecell[c]{$\mathcal{G}= (\mathcal{V}, \mathcal{E})$\\ $ \mathcal{E} = \mathcal{E}_c \cup \mathcal{E}_e$ \\$ \mathcal{V} = \mathcal{V}_{\mathcal{T}} \cup \mathcal{V}_{\mathcal{K}} \cup \mathcal{V}_e$} 
    & \makecell[l]{Edges set $\mathcal{E}_c \subseteq \mathcal{V}_{\mathcal{T}} \times \mathcal{V}_{\mathcal{K}}$ \\ Edges set $\mathcal{E}_e \subseteq \mathcal{V}_{e} \times \mathcal{V}_{\mathcal{K}}$ \\ Entity nodes set $\mathcal{V}_{e}$ \\ Chunk nodes set $\mathcal{V}_{\mathcal{T}}$  \\ Knowledge unit nodes set $\mathcal{V}_{\mathcal{K}}$} \\
    \hline
  \end{tabular}
\end{table}

\begin{problem}[Graph-based RAG]
\label{prob:graph-rag}
Given an LLM, a text corpora $\mathcal T$, and a set of questions $Q$, find the relevant information for $Q$ from the KG built on $\mathcal T$, which can be used to augment $Q$ so that the LLM can generate high-quality answers for $Q$.
\end{problem}

\section{Our Proposed Approach: Clue-RAG}
\label{sec:clue-rag}

In this section, we develop a novel graph-based RAG, called Clue-RAG, which addresses the two major limitations in existing methods by manipulating 
\textit{text \underline{C}hunk},
\textit{know\underline{l}edge \underline{u}nit}, and
\textit{\underline{e}ntity}.
Specifically, we design a multi-partite graph index that supports semantically coherent retrieval across multiple levels of granularity, coupled with a hybrid strategy for knowledge unit extraction that balances LLM utility with computational cost.
Besides, we propose \texttt{Q-Iter}, an iterative retrieval strategy that improves retrieval accuracy.
We first provide an overview of Clue-RAG in Section~\ref{overview} and then describe its two core modules in Sections~\ref{Clue-index} and \ref{Q-Iter}.

\subsection{Overview}
\label{overview}

At a high level, Clue-RAG consists of two modules: offline indexing \texttt{Clue-Index} and online retrieval \texttt{Q-Iter}.
As shown in Figure \ref{fig:overview}, the offline \texttt{Clue-Index} has two stages:
\begin{itemize}[leftmargin=*]
    \item {\bf Stage-1: hybrid extraction.} Given a text corpora \(\mathcal{T}\) and a token constraint coefficient \(\alpha \in [0,1]\), a subset of core text chunks \(\mathcal{T}_c \subseteq \mathcal{T} \)  is selected based on relevance metrics. For each \(t \in \mathcal{T}_c\), an LLM is employed to extract disambiguated knowledge units. For \(t \in \mathcal{T} \setminus \mathcal{T}_c\), we use an NLP tool to segment the text into sentence-level units. Finally, entities are extracted from knowledge units using an NLP tool.
    \item {\bf Stage-2: graph construction.} We construct a multi-partite graph index $\mathcal{G} = (\mathcal{V}, \mathcal{E})$, including three types of nodes: chunk, knowledge unit, and entity.
    In $\mathcal{G}$, edges only connect nodes of different types: specifically, chunk nodes are linked to their corresponding knowledge unit nodes, and knowledge unit nodes are further connected to the entities they reference.
\end{itemize}

During the online retrieval, \texttt{Q-Iter} begins by extracting entities from the input query $q$ to activate related entities, while semantic search anchors initial knowledge units. These units expand the entity set via referenced entities, enabling constrained multihop traversal over $\mathcal{G}$ to retrieve additional relevant units. The resulting units are mapped to their source chunks and re-ranked to form the final context for answer generation.

\begin{algorithm}[htb]
\small
\caption{\ttfamily{Clue-Index ($\mathcal{T}, \alpha$)}}
\label{alg:graph-index}

\textbf{Input:} Text corpora $\mathcal{T} = \{t_1, \dots, t_n\}$, constraint $\alpha \in [0,1]$
 
\textbf{Output:} A multi-partite graph index $\mathcal{G} = (\mathcal{V}, \mathcal{E})$

$\mathcal{T}_c \gets$ \ttfamily{Chunk-Selection ($\mathcal{T}, \alpha$)} in Algorithm \ref{alg:chunk_selection};
 
$\mathcal{V}_{\mathcal{T}} \gets \{t_i\mid t_i \in \mathcal{T}\}$, $\mathcal{V}_{\mathcal{K}} \gets \emptyset$, $\mathcal{V}_{e} \gets \emptyset$; 

$\mathcal{E}_c \gets \emptyset$, $\mathcal{E}_e \gets \emptyset$; 

\ForEach{$t_i \in \mathcal{V}_{\mathcal{T}}$} {
    \If{$t_i \in \mathcal{T}_c$} {
        $\mathcal{K}_i \gets \text{extracted by LLM from } t_i$\;
    }
    \Else {
        $\mathcal{K}_i \gets \text{split by NLP tool from } t_i$\;
    }

    $\mathcal{V}_{\mathcal{K}} \gets \mathcal{V}_{\mathcal{K}} \cup \mathcal{K}_i$; $\mathcal{E}_c \gets \mathcal{E}_c \cup \{(t_i, k) \mid k \in \mathcal{K}_i\}$; 
}

\ForEach {$k \in \mathcal{V}_{\mathcal{K}}$} {
         
        $\mathcal{V}_k \gets \text{NER}(k)$; $\mathcal{V}_e \gets \mathcal{V}_e \cup \mathcal{V}_k$; 

        $\mathcal{E}_e \gets \mathcal{E}_e \cup \{(v, k) \mid v \in \mathcal{V}_k\}$; 
    }

 \Return $\mathcal{G} = (\mathcal{V}_{\mathcal{T}} \cup \mathcal{V}_{\mathcal{K}} \cup \mathcal{V}_e, \mathcal{E}_c \cup \mathcal{E}_e)$;

\end{algorithm}


\begin{algorithm}[htb]
\small
\caption{\ttfamily{Chunk-Selection ($\mathcal{T}, \alpha$)}}
\label{alg:chunk_selection}

 \textbf{Input:} Text corpora $\mathcal{T} = \{t_1, \dots, t_n\}$, constraint $\alpha \in [0,1]$
 
 \textbf{Output:} a core subset of text chunks \(\mathcal{T}_c \in \mathcal{T}\)

 $W_{\text{total}} \gets 0$, $\mathcal{S} \gets \emptyset$;
 
\ForEach{$t_i \in \mathcal{T}$} {
     $w_i \gets \text{TokenLength}(t_i)$;
     
     $v_i \gets \text{BLEU}(t_i, \mathcal{T} \setminus \{t_i\})$; \Comment{Standard BLEU calculation}
     
     $W_{\text{total}} \gets W_{\text{total}} + w_i$; $\mathcal{S} \gets \mathcal{S} \cup \{(w_i, v_i)\}$;
}

 $W_{\text{max}} \gets \lceil \alpha \cdot W_{\text{total}} \rceil$;
 
 $\mathcal{T}_c \gets \text{Knapsack}(\mathcal{S}, W_{\text{max}})$; 
 
 \Return $\mathcal{T}_c$;

\end{algorithm}

\subsection{Offline index: {\ttfamily Clue-Index}}
\label{Clue-index}

We illustrate the two key stages of indexing as follows.

\noindent \textbf{Stage-1: hybrid extraction.}
As shown in Algorithm \ref{alg:graph-index}, the hybrid extraction process begins by selecting a subset of core text chunks \(\mathcal{T}_c \subseteq \mathcal{T}\) using Algorithm \ref{alg:chunk_selection}, based on a token constraint coefficient \(\alpha \in [0,1]\).
In Algorithm \ref{alg:chunk_selection}, we formulate the selection stage as a 0-1 knapsack problem, where each chunk’s relevance score determines its assigned value (line 5) and its token length corresponds to its weight (line 6), with the goal of selecting subsets to maximize total relevance under the token budget (line 9) — a classic optimization paradigm identical to 0-1 knapsack problem \cite{sahni1975approximate}.


In Algorithm \ref{alg:graph-index}, once \(\mathcal{T}_c\) is identified (line 3), each text chunk \(t_i\) is processed according to its selection status (lines 6–11). For \(t_i \in \mathcal{T}_c\), an LLM extracts disambiguated knowledge units \(\mathcal{K}_i\) that capture core semantics.
For \(t_i \in \mathcal{T} \setminus \mathcal{T}_c\), a lightweight NLP tool \cite{Bird_Natural_Language_Processing_2009} is used to segment the text into sentences as knowledge units \(\mathcal{K}_i\) to ensure broad coverage at lower computational cost.
All text chunks and extracted knowledge units are instantiated as text nodes $\mathcal{V}_\mathcal{T}$ and knowledge unit nodes $\mathcal{V}_\mathcal{K}$, respectively, forming the upper and intermediate layers of the multi-partite graph.
Finally, the Named Entity Recognition (NER) component in spaCy \cite{Honnibal_spaCy_Industrial-strength_Natural_2020} is utilized to extract entities \(\mathcal{V}_k\) for each knowledge unit \(k\) (lines 12-13).
All these extracted entities \(\mathcal{V}_k\) are collected and form entity nodes  \(\mathcal{V}_e\), acting as anchors for precise information retrieval.

\noindent \textbf{Stage-2: graph construction.}
As shown in Figure~\ref{fig:overview}, {\ttfamily Clue-Index} is a multi-partite graph \(\mathcal{G}\) built from a text corpora \(\mathcal{T}\), organizing information across three semantic levels.
In the upper layer, chunk nodes \(\mathcal{V}_\mathcal{T}\) represent coarse-grained content by encoding text chunks. The intermediate layer contains knowledge unit nodes \(\mathcal{V}_\mathcal{K}\), which capture semantically richer information distilled from the text.
In the lower layer, entity nodes \(\mathcal{V}_e\) encode fine-grained factual elements extracted from knowledge units. The graph is then connected by two types of edges: \(\mathcal{E}_c\), linking each chunk node to its corresponding knowledge units, and \(\mathcal{E}_e\), connecting knowledge units to the entities they mention.
We present the construction algorithm in Algorithm \ref{alg:graph-index}.

\subsection{Online retrieval algorithm: {\ttfamily Q-Iter}}
\label{Q-Iter}

Given a query $q$ and a multi-partite graph index $\mathcal{G}$, our retrieval algorithm {\ttfamily Q-Iter}, outlined in Algorithm \ref{alg:reasoning}, identifies relevant information of $q$ from the corpora.
It has four input parameters, i.e., the number of most relevant results $K$, maximum search depth $D$, beam size $M$ per depth level, and returned chunks size $N$.
{\ttfamily Q-Iter} proceeds through three carefully designed steps as follows:

\noindent \textbf{Step-1: entity anchoring.}
The goal is to identify query-relevant seed nodes \(\mathcal{V}_q^{(0)}\).
Inspired by cognitive theories of spreading activation \cite{anderson1983spreading,collins1975spreading}, we first extract entities from the input query using an LLM and identify their top-$K$ most similar entity nodes in \(\mathcal{G}\) to form the initial query set \(\mathcal{V}_q^{(0)}\). 
In parallel, we retrieve the top-$K$ knowledge units that are most semantically similar to the query and expand \(\mathcal{V}_q^{(0)}\) by including all entities referenced in these units.
This dual expansion incorporates both directly matched entities and entities introduced via semantically aligned knowledge units, enhancing the coverage and specificity of subsequent multihop retrieval.
We refer to this combined process of Spreading Activation and Knowledge Anchoring as {\ttfamily Entity Anchoring} shown in Algorithm \ref{alg:entity_anchor}, which yields the initial seed nodes \(\mathcal{V}_q^{(0)}\).

\begin{algorithm}[htbp]
\small
\caption{\ttfamily{Q-Iter($q, \mathcal{G}, K, D, M, N$)}}
\label{alg:reasoning}

 \textbf{Input:} \(q\), $\mathcal{G} = (\mathcal{V}_{\mathcal{T}} \cup \mathcal{V}_{\mathcal{K}} \cup \mathcal{V}_e, \mathcal{E}_c \cup \mathcal{E}_e)$, \(K\), \(D\), \(M\), \(N\)
 
 \textbf{Output:} \text{top-}\({N}\) relevant text chunks \(\mathcal{T}^{*}\)

$\mathcal{V}_q^{(0)} \gets$ {\ttfamily Entity Anchoring}$(q, K, \mathcal{G})$ in Algorithm \ref{alg:entity_anchor};

 \textcolor{gray}{\# Iterative Retrieval}

 $\mathcal{Q}^{(0)} \gets \bigl\{(v, \bm{\phi}(q),\emptyset) \mid v \in \mathcal{V}_q^{(0)}\bigr\}$ ; \( \mathcal{C^{*}} \gets \emptyset\);

\ForEach{\(d=1,\dots,D\)} {
  \(\mathcal{C} \gets \emptyset\);
  
  \ForEach{\((v,\boldsymbol{\phi}_v,\mathcal{S}_v)\in\mathcal{Q}^{(d-1)}\)}{

    $\mathcal{K}_v \gets \mathop{\arg\max}\limits_{\substack{k \in \mathcal{V}_{\mathcal{K}} \setminus \mathcal{S}_v \land (v,k) \in \mathcal{E}_e \land |\mathcal{K}_v| = K}} \cos\bigl(\bm{\phi}(k), \bm{\phi}_v\bigr)$; 
    
    \ForEach{\(k\in\mathcal{K}_v\)}{
      \(\boldsymbol{\phi}^{\prime}_v \gets \boldsymbol{\phi}_v - \boldsymbol{\phi}(k)\); \Comment{Update query embedding}
      
      \(\mathcal{S}_v^{\prime}\gets \text{sort} (\mathcal{S}_v\cup\{k\})\); \Comment{Dictionary sorting}
       
       \( score \gets \mathcal{R} (q, \oplus (\mathcal{S}^{\prime}_v))\); \Comment{Re-rank} 
      
      \(\mathcal{V}^{\prime}\gets \{v\mid (v,k)\in\mathcal{E}_e\}\);
      
      \( \mathcal{C} \gets  \mathcal{C} \cup \{(\mathcal{V}^{\prime},\boldsymbol{\phi}_v^{\prime},\mathcal{S}_v^{\prime}, score)  \}\);
        }
    }

  \(\mathcal{C}^{\prime} \gets \mathop{\arg\max} \limits_{\substack{(\mathcal{V},\boldsymbol{\phi},\mathcal{S}, score) \in \mathcal{C} \land \vert \mathcal{C}^{\prime} \vert = M}} score\); \Comment{Pruning}

  \( \mathcal{C^{*}} \gets \mathcal{C^{*}} \cup \mathcal{C}^{\prime} \);

  $\mathcal{Q}^{(d)} \gets \bigcup_{(\mathcal{V},\boldsymbol{\phi},\mathcal{S}, score)\in \mathcal{C}^{\prime}} \{ (v, \boldsymbol{\phi},\mathcal{S}) \mid v \in \mathcal{V} \}$;
}

 \textcolor{gray}{\# Final Re-ranking}
 
 \(\mathcal{T}^{\prime} \gets \emptyset\);

\ForEach{\((\mathcal{V},\boldsymbol{\phi},\mathcal{S}, score) \in \mathcal{C^{*}}\)}{
  $\mathcal{T}^{\prime} \gets \mathcal{T}^{\prime} \cup \{ t \mid k \in \mathcal{S} \land (t,k) \in \mathcal{E}_c \}$;
}

 \(\mathcal{T}^{*} \gets \mathop{\arg\max} \limits_{\substack{t \in \mathcal{T}^{\prime} \land |\mathcal{T}^{*}|=N}} \mathcal{R}(q, t)\);

 \Return $\mathcal{T}^{*}$;

\end{algorithm}

\begin{figure*}[htbp]
    \centering
    \includegraphics[width=\textwidth]{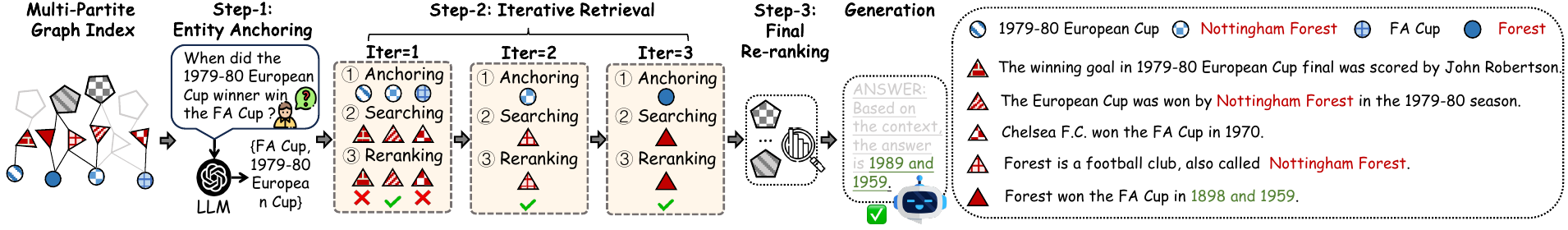}
    \caption{A toy example illustrating the online retrieval.}
    \label{fig:top_example}
\end{figure*}

\noindent \textbf{Step-2: iterative retrieval.}
The key idea is to progressively gather highly query-relevant and non-redundant knowledge units at each depth.
Given the seed nodes $\mathcal{V}_q^{(0)}$, Algorithm \ref{alg:reasoning} first initializes $\mathcal{Q}^{(0)}$ with tuples $(v, \bm{\phi}(q), \emptyset)$ for each anchor $v \in \mathcal{V}_q^{(0)}$ (line 5), where $\bm{\phi}(q)$ is the original query embedding that guides the semantic search and $\emptyset$ will accumulate the retrieved knowledge units.
For each depth $d$ from $1$ to $D$, it processes the query queue $\mathcal{Q}^{(d-1)}$ (lines 6-18). 
For each $(v, \bm{\phi}_v, \mathcal{S}_v) \in \mathcal{Q}^{(d-1)}$, it retrieves top-$K$ new knowledge units $\mathcal{K}_v \subseteq  \mathcal{V}_{\mathcal{K}}\setminus\mathcal{S}_v$ using $\bm{\phi}_v$, ensuring relevance while avoiding redundancy (lines 8-9).
For each $k \in \mathcal{K}_v$, it first updates the query embedding by subtracting $k$'s embedding (lines 10-15), dynamically shifting focus to uncovered information and avoiding redundancy (This mechanism is termed Query Updating, as demonstrated in Experiment \ref{exp:ablation}.).
Next, the algorithm adds $k$ to $\mathcal{S}_v$, sorts it canonically, then uses re-ranker $\mathcal{R}$ to score $\oplus(\mathcal{S}_v^{\prime})$ for coherence with $q$.
Unlike semantic similarity measures, $\mathcal{R}$ specializes in query-context relevance scoring, allowing a more accurate assessment of knowledge units' relevance to $q$.
Then, the algorithm: (1) finds adjacent entities $\mathcal{V}^{\prime}$ via $k$'s edges as next-depth anchors, and (2) updates $\mathcal{C}$ with the new state tuple and its re-ranked $score$.
Finally, the algorithm prunes $\mathcal{C}$ to the top-$M$ tuples $\mathcal{C}^{\prime}$ based on their re-ranked $scores$, merges them into $\mathcal{C}^{*}$, and initializes $\mathcal{Q}^{(d)}$ for the next iteration (lines 16-18), where the re-ranker $\mathcal{R}$ guarantees $\mathcal{C}^{*}$ contain maximal evidential support for answering $q$ while maintaining computational efficiency.

\noindent \textbf{Step-3: final re-ranking.}
The key idea is to retrieve the most query-relevant text chunks from $\mathcal{C}^{*}$ via re-ranking for augmented generation.
The algorithm first retrieves text nodes $\mathcal{T}^{\prime}$ citing $\mathcal{S}$ via $\mathcal{E}_c$, and then selects top-$N$ by their re-ranked $scores$ to form final chunks $\mathcal{T}^{*}$ for augmented generation (lines 19-23).

\begin{example}
Given a query $q=$ ``When did the 1979-80 European Cup winner win the FA Cup?'', then {\ttfamily Q-Iter} executes the three steps:

\noindent \textbf{Step-1 (entity anchoring):}
The LLM first extracts entities $\mathcal{V}_q^{(0)}$ = \{\textit{1979-80 European Cup}, \textit{FA Cup}\} from $q$. Simultaneously, it retrieves the most semantically similar knowledge unit \textit{``The European Cup was won by Nottingham Forest in the 1979-80 season''} using the query embedding, expanding $\mathcal{V}_q^{(0)}$ to \{\textit{1979-80 European Cup}, \textit{FA Cup}, \textit{Nottingham Forest}\}.  

\noindent \textbf{Step-2 (iterative retrieval):}
In the first iteration, the anchoring stage initializes with the seed nodes from Step 1.
For each entity in $\mathcal{V}_q^{(0)}$, the algorithm retrieves the top-$K$ ($K=1$) semantically similar knowledge unit linked to it via the query embedding.
For instance, \textit{FA Cup} retrieves \textit{Chelsea F.C. won the FA Cup in 1970}.
After each iteration, Step 2 applies re-ranker $\mathcal{R}$ to score knowledge units against $q$.
In this toy example, only the top-$M$ ($M=1$) relevant units are retained per iteration.
The process terminates at maximum depth $D=3$. The complete retrieval path originates from \textit{Nottingham Forest} to \textit{Forest won the FA Cup in 1898 and 1959}.

\noindent \textbf{Step-3 (final re-ranking):} The algorithm maps the retrieved knowledge units back to their source text chunks, then re-ranks them based on $q$, and finally uses the most relevant chunks for RAG.

\begin{algorithm}[htbp]
\small
\caption{\ttfamily{Entity Anchoring($q, \mathcal{G}, K$)}}
\label{alg:entity_anchor}

 \textbf{Input:} \(q\), \(K\), $\mathcal{G} = (\mathcal{V}_{\mathcal{T}} \cup \mathcal{V}_{\mathcal{K}} \cup \mathcal{V}_e, \mathcal{E}_c \cup \mathcal{E}_e)$
 
 \textbf{Output:} initial seed nodes $\mathcal{V}_q^{(0)}$

 \textcolor{gray}{\# Spreading Activation}
 
 \(\mathcal{V}_q \gets \mathrm{NER}(q)\); \(\mathcal{V}_q^{(0)} \gets \emptyset \);
 
\ForEach{\(v\in\mathcal{V}_q\)} {
  $\mathcal{N}_v \gets \mathop{\arg\max} \limits_{\substack{v^{\prime} \in \mathcal{V}_e} \land |\mathcal{N}_v| = K} \cos\bigl(\bm{\phi}(v^{\prime}),\,\bm{\phi}(v)\bigr)$;
  
  \(\mathcal{V}_q^{(0)} \gets \mathcal{V}_q^{(0)} \cup \mathcal{N}_v\);
}

 \textcolor{gray}{\# Knowledge Anchoring}

 $\mathcal{K}_q \gets \mathop{\arg\max}\limits_{\substack{k \in \mathcal{V}_\mathcal{K} \land |\mathcal{K}_q| = K}} \cos\bigl(\bm{\phi}(k), \bm{\phi}(q)\bigr)$;

 $\mathcal{V}_q^{(0)} \gets \mathcal{V}_q^{(0)} \cup \bigcup_{k \in \mathcal{K}_q} \{ v \mid (v,k) \in \mathcal{E}_e \}$;

 \Return $\mathcal{V}_q^{(0)}$

\end{algorithm}

\end{example}

\section{Experiments}
\label{sec:experiments}


In this section, we assess the effectiveness of Clue-RAG by addressing the following research questions:

\begin{itemize}[leftmargin=*]
\item \textbf{Q1}: How does Clue-RAG compare to existing baselines in terms of QA performance and cost efficiency?

\item \textbf{Q2}: How does our hybrid extraction strategy compare to alternatives and full LLM-based extraction?

\item \textbf{Q3}: What are the contributions of each step of {\ttfamily Q-Iter}?

\item \textbf{Q4}: How sensitive is Clue-RAG to hyperparameter settings?
\end{itemize}


\subsection{Experimental Setup}

\begin{table*}[htbp]
\centering
\small
\caption{Overall performance of RAG solutions. The best and second-best results among the ten competitor solutions (excluding the Clue-RAG series) in each column are highlighted in bold and underlined, respectively.}
\small
\renewcommand{\arraystretch}{1.00}
\begin{tabular}{lcccccccccccccc}
\toprule
& \multicolumn{6}{c}{LLaMA3.0-8B} & \multicolumn{6}{c}{Qwen2.5-32B} & \multicolumn{2}{c}{}\\
\cmidrule(lr){2-7} \cmidrule(lr){8-13} 
 & \multicolumn{2}{c}{MuSiQue} & \multicolumn{2}{c}{HotpotQA} & \multicolumn{2}{c}{2Wiki} & \multicolumn{2}{c}{MuSiQue} & \multicolumn{2}{c}{HotpotQA} & \multicolumn{2}{c}{2Wiki}  & \multicolumn{2}{c}{Avg}\\ 
\cmidrule(lr){2-3} \cmidrule(lr){4-5} \cmidrule(lr){6-7} \cmidrule(lr){8-9} \cmidrule(lr){10-11} \cmidrule(lr){12-13} \cmidrule(lr){14-15}
Method & F1 & Acc. & F1 & Acc. & F1 & Acc. & F1 & Acc. & F1 & Acc. & F1 & Acc. & F1 & Acc. \\ 
\midrule
\rowcolor{gray!10} \multicolumn{15}{c}{\bfseries Simple Baselines} \\ 
{\tt Zero-shot} & 7.14 & 5.11 & 24.34 & 25.30 & 18.88 & 33.78 & 9.62 & 5.71 & 25.51 & 25.90 & 26.44 & 27.10 & 18.66 & 20.48 \\
{\tt Vanilla} & 12.28 & 14.21 & 40.62 & 52.50 & 18.22 & 34.00 & 20.97 & 21.02 & 53.55 & 60.50 & 30.94 & 44.80 & 29.43 & 37.84 \\
\rowcolor{gray!10} \multicolumn{15}{c}{\bfseries Graph-Structured Baselines} \\ 
{\tt LGraphRAG} & 9.23 & 14.21 & 24.81 & 43.90 & 15.11 & 42.20 & 13.12 & 19.12 & 34.04 & 53.20 & 21.57 & 46.20 & 19.65 & 36.47 \\
{\tt LLightRAG} & 7.94 & 6.91 & 25.43 & 36.30 & 17.35 & 33.60 & 13.82 & 18.52 & 34.09 & 49.10 & 22.53 & 48.40 & 20.19 & 32.14 \\
{\tt GLightRAG} & 5.34 & 5.14 & 17.80 & 27.40 & 11.82 & 23.20 & 9.31 & 11.51 & 21.69 & 34.30 & 8.09 & 35.70 & 12.34 & 22.88 \\
{\tt HLightRAG} & 7.87 & 9.41 & 25.82 & 37.40 & 13.16 & 28.40 & 14.57 & 22.62 & 35.89 & 55.60 & 19.16 & 52.30 & 19.41 & 34.29 \\
{\tt HippoRAG} & 11.58 & 11.81 & 39.63 & 44.40 & 24.94 & \underline{48.20} & 15.00 & 20.42 & 40.36 & 52.20 & 26.95 & 53.70 & 26.41 & 38.46 \\
{\tt KETRAG} & \underline{16.25} & 12.51 & \underline{48.75} & 47.10 & \underline{27.90} & 41.40 & 21.08 & 16.52 & \underline{61.94} & 57.70 & \underline{50.02} & 49.90 & \underline{37.66} & 37.52 \\
\rowcolor{gray!10} \multicolumn{15}{c}{\bfseries Tree-Structured Baselines} \\ 
{\tt RAPTOR} & 10.58 & 16.22 & 34.18 & \underline{56.50} & 16.92 & 43.30 & 17.31 & 21.12 & 47.09 & 63.70 & 28.88 & 47.20 & 25.83 & 41.34 \\
{\tt SIRERAG} & 11.99 & \underline{21.22} & 34.02 & 55.70 & 18.34 & 43.90 & \underline{22.46} & \underline{31.93} & 46.75 & \textbf{65.10} & 33.88 & \underline{57.10} & 27.91 & \underline{45.83} \\
\midrule
{\tt Clue-RAG-0.0} & 24.10 & 21.02 & 53.18 & 57.20 & 32.19 & 46.60 & 34.26 & 31.33 & 62.14 & 63.50 & 49.47 & 54.70 & 42.56 & 45.73 \\
{\tt Clue-RAG-0.5} & \textbf{25.60} & 22.12 & 55.80 & \textbf{59.50} & 35.52 & 48.90 & \textbf{36.68} & \textbf{32.93} & 63.44 & 64.10 & 52.49 & 58.80 & 44.92 & 47.73 \\

{\tt Clue-RAG-1.0} & 25.48 & \textbf{22.32} & \textbf{55.97} & 59.20 & \textbf{37.41} & \textbf{50.70} & 36.50 & 32.83 & \textbf{64.02} & \underline{64.70} & \textbf{55.20} & \textbf{61.00} & \textbf{45.76} 
& \textbf{48.46} \\


\bottomrule
\end{tabular}
\label{tab:main_result}
\end{table*}

\noindent \textbf{Datasets.}
We evaluate Clue-RAG on three multihop QA benchmarks: MuSiQue \cite{trivedi2022musique}, HotpotQA \cite{yang2018HotpotQA}, and 2WikiMultiHopQA (2Wiki Below) \cite{ho2020constructing}. Following prior work \cite{gutierrez2024hipporag, press2022measuring, trivedi2022interleaving, zhang2024sirerag}, we use 1,000 validation questions per dataset, with corresponding paragraphs preprocessed into $\mathcal{T}$. See Appendix Table \ref{tab:dataset} for statistics.


\noindent \textbf{Metrics.}
We evaluate each solution's performance in terms of QA performance and cost efficiency.
For QA performance, we report Accuracy (Acc. below) and F1 score (F1 below), where the former one measures whether the golden answers are included in the generation answer rather than requiring exact match \cite{asai2023self, mallen2022not}, and the latter one is computed via token-level overlap between golden and generated answers, balancing precision and recall to assess both answer completeness and correctness \cite{gutierrez2024hipporag, zhang2024sirerag}.
For cost efficiency, we quantify two metrics: total token expenditure for offline indexing and average token consumption per online query $q$.

\noindent \textbf{Baselines.} We test the following 13 solutions in 4 categories.

\begin{itemize}[leftmargin=*]
    \item \textbf{Simple baselines:} We consider 2 straightforward approaches: (1) the Zero-shot method, which directly applies an LLM for QA without any retrieval, and (2) VanillaRAG, which retrieves relevant passages through query embedding similarity before using these passages as context for generation.
    
    \item \textbf{Graph-based baselines:} We evaluate 6 graph-based baselines that construct text-attribute KG using LLMs for content retrieval: HippoRAG \cite{gutierrez2024hipporag}, KETRAG \cite{huang2025ket}, and local search of GraphRAG (LGraphRAG) \cite{edge2024local}, and 3 versions of LightRAG (LLightRAG for local search, GLightRAG for global search, and HLightRAG for hybrid search) \cite{guo2024lightrag}, excluding GraphRAG's global version as it targets abstract QA rather than multi-hop QA.
    
    \item \textbf{Tree-based baselines:} We evaluate 2 tree-based baselines: RAPTOR \cite{sarthi2024raptor} and SIRERAG \cite{zhang2024sirerag}, both constructing multi-level trees for passage organization and retrieval.

     \item \textbf{Our proposed solutions:} Our approach supports three operational modes, determined by a hyperparameter \(\alpha \in [0,1]\), which controls the proportion of tokens in \(\mathcal{T}\) processed by LLMs during offline indexing.
     Clue-RAG-0.0 (\(\alpha=0.0\)) demonstrates the strong performance of our LLM-free graph index and retrieval strategy.
     Clue-RAG-0.5 (\(\alpha=0.5\)) highlights the effectiveness of our hybrid extraction strategy, achieving strong performance while reducing 50\% indexing cost.
     Clue-RAG-1.0 (\(\alpha=1.0\)) represents the full-capacity configuration, where all text chunks are processed by LLMs, enabling the system to achieve SOTA performance.
\end{itemize}

\noindent \textbf{Settings.} All experiments are conducted on a Linux machine with an Intel Xeon 2.0 GHz CPU, 1024 GB of memory, and 8 NVIDIA GeForce RTX A5000 GPUs (each with 24 GB of memory), capable of running both LLaMA3.0-8B \cite{grattafiori2024llama} and Qwen2.5-32B \cite{bai2023qwen} LLM models. To ensure a fair comparison, all methods are implemented under a unified framework \cite{zhou2025depth} using their default configurations. Additional experimental details are provided in Appendix \ref{Experimental_Settings}.
\subsection{Overall performance evaluation (Q1)}
\label{exp:main}

\begin{table*}[htbp]
\centering
\small
\caption{Performance comparison of Clue-RAG-0.5 using different core chunk selection strategies. The table highlights the best and second-best results in each column with bold and underlined formatting, respectively.}
\renewcommand{\arraystretch}{1.0}
\begin{tabular}{lcccccccccccccc}
\toprule
& \multicolumn{6}{c}{LLaMA3.0-8B} & \multicolumn{6}{c}{Qwen2.5-32B} & \multicolumn{2}{c}{}\\
\cmidrule(lr){2-7} \cmidrule(lr){8-13} 
 & \multicolumn{2}{c}{MuSiQue} & \multicolumn{2}{c}{HotpotQA} & \multicolumn{2}{c}{2Wiki} & \multicolumn{2}{c}{MuSiQue} & \multicolumn{2}{c}{HotpotQA} & \multicolumn{2}{c}{2Wiki}  & \multicolumn{2}{c}{Avg}\\ 
\cmidrule(lr){2-3} \cmidrule(lr){4-5} \cmidrule(lr){6-7} \cmidrule(lr){8-9} \cmidrule(lr){10-11} \cmidrule(lr){12-13} \cmidrule(lr){14-15}
Method & F1 & Acc. & F1 & Acc. & F1 & Acc. & F1 & Acc. & F1 & Acc. & F1 & Acc. & F1 & Acc. \\ 
\midrule

{\tt Clue-RAG-0.5 (RANDOM)} & \underline{25.18} & 21.32 & 54.44 & 57.60 & 34.00 & 46.90 & \underline{36.61} & \textbf{33.93} & 61.59 & 63.40 & 49.49 & 54.70 & 43.55 & 46.31 \\

{\tt Clue-RAG-0.5 (COS)} & 24.64 & \underline{21.42} & \underline{55.80} & \underline{59.10} & \underline{35.18} & \textbf{49.10} & 36.19 & 32.13 & \textbf{63.65} & \textbf{64.90} & \textbf{52.83} & \textbf{59.10} & \underline{44.72} & \underline{47.63} \\

{\tt Clue-RAG-0.5 (BLEU)} & \textbf{25.60} & \textbf{22.12} & \underline{55.80} & \textbf{59.50} & \textbf{35.52} & \underline{48.90} & \textbf{36.68} & \underline{32.93} & \underline{63.44} & \underline{64.10} & \underline{52.49} &\underline{ 58.80} & \textbf{44.92} & \textbf{47.73} \\

\midrule

{\tt Clue-RAG-1.0 (BLEU)} & 25.48 & 22.32 & 55.97 & 59.20 & 37.41 & 50.70 & 36.50 & 32.83 & 64.02 & 64.70 & 55.20 & 61.00 & 45.76 & 48.46 \\

\bottomrule
\end{tabular}
\label{tab:selection}
\end{table*}





In the first set of experiments, we compare Clue-RAG against ten existing solutions in terms of QA performance and cost efficiency.
As shown in Table \ref{tab:main_result}, Clue-RAG consistently achieves superior QA performance across multiple benchmark datasets when evaluated with different LLMs, with Clue-RAG-1.0 establishing SOTA performance in both average F1 and Acc.
Specifically, it outperforms the strong baseline KETRAG by 21.53\% in average F1, while also surpassing another competitive baseline SIRERAG by 5.75\% in average Acc.
Besides, Clue-RAG-0.5 demonstrates competitive QA performance relative to KETRAG, and achieves improvements of 19.29\%/27.19\% in average F1/Acc.
Notably, Clue-RAG-0.0, which entirely avoids LLM usage in indexing, still matches or exceeds the capabilities of two strong baselines, achieving a 13.01\%/52.50\% average F1 improvement over KETRAG and SIRERAG, respectively.
Beyond outperforming these two competitive baselines, Clue-RAG demonstrates superior QA performance across all evaluated datasets compared to the other eight alternatives. This consistent advantage highlights the effectiveness of its innovative indexing and retrieval strategies in multi-hop QA scenery.

Meanwhile, compared to KETRAG, Clue-RAG-0.5 reduces token costs by up to 9.41\%/83.87\% during offline indexing and online retrieval, as shown in Figure \ref{fig:token_compare_optimized} and \ref{fig:query_token}.
Remarkably, during offline indexing, Clue-RAG-0.0 reduces token costs by 100\% compared to KETRAG and SIRERAG. During online retrieval, it reduces up to 82.92\%/56.06\% token costs compared to these two strong baselines.
Hence, Clue-RAG is token-cost efficient.

\definecolor{c0}{HTML}{642915}  
\definecolor{c1}{HTML}{963e20}  
\definecolor{c2}{HTML}{c7522a}  
\definecolor{c3}{HTML}{d68a58}  
\definecolor{c4}{HTML}{e5c185}  
\definecolor{c5}{HTML}{fbf2c4}  
\definecolor{c6}{HTML}{d4e8b7}  
\definecolor{c7}{HTML}{b8cdab}  
\definecolor{c8}{HTML}{74a892}  
\definecolor{c9}{HTML}{3a978c}  
\definecolor{c10}{HTML}{008585} 
\definecolor{c11}{HTML}{006b6b} 
\definecolor{c12}{HTML}{004f4f} 


\begin{figure}[htbp]
    \centering
    \setlength{\abovecaptionskip}{2pt}
    \setlength{\belowcaptionskip}{-5pt}
    \begin{tikzpicture}[scale=0.52]
        \begin{axis}[
            ybar=0.0cm,
            bar width=0.45cm,
            width=0.9\textwidth,
            height=0.35\textwidth,
            xtick={1,2,3},
            xticklabels={MuSiQue, HotpotQA, 2Wiki},
            xticklabel style={font=\Large},
            yticklabel style={font=\large},
            legend style={
                at={(0.5,1.25)},  
                anchor=north,
                legend columns=5,  
                draw=none,
                font=\large,
                cells={anchor=west},
                column sep=4pt,    
                row sep=2pt,
            },
            legend image code/.code={
                \draw [#1] (0cm,-0.1cm) rectangle (0.3cm,0.1cm);
            },
            xmin=0.5, xmax=3.5,
            ymin=1e6, ymax=1.2e8,
            ymode=log,
            ytick={1e6,1e7,1e8},
            yticklabels={$10^6$,$10^7$,$10^8$},
            ylabel={\Large Tokens (log scale)},
            axis line style={line width=0.5pt},
            tick style={line width=1.0pt},
            xtick style={draw=none},
        ]
        
        \addplot[fill=c2, draw=black,line width=0.5pt] coordinates {(1,100886430) (2,87048145) (3,51458539)}; 
        \addplot[fill=c4, draw=black,line width=0.5pt] coordinates {(1,71928913) (2,61390248) (3,37175723)}; 
        \addplot[fill=c6, draw=black,line width=0.5pt] coordinates {(1,15535482) (2,13749587) (3,7932348)};  
        \addplot[fill=c7, draw=black,line width=0.5pt] coordinates {(1,6998226) (2,6181250) (3,3500674)};  
        \addplot[fill=c8, draw=black,line width=0.5pt] coordinates {(1,3966996) (2,3624612) (3,2108092)};  
        \addplot[fill=c9, draw=black,line width=0.5pt] coordinates {(1,23215958) (2,20885165) (3,12157786)}; 
        \addplot[fill=c10, draw=black,line width=0.5pt] coordinates {(1,1) (2,1) (3,1)};         
        \addplot[fill=c11, draw=black,line width=0.5pt] coordinates {(1,6460942) (2,5599630.5) (3,3357518)};   
        \addplot[fill=c12, draw=black,line width=0.5pt] coordinates {(1,12921884) (2,11199261) (3,6715036)};  

        \legend{
            GraphRAG,
            LightRAG,
            HippoRAG,
            KETRAG,
            RAPTOR,
            SIRERAG,
            Clue-RAG-0.0,
            Clue-RAG-0.5,
            Clue-RAG-1.0,
        }
        \end{axis}
    \end{tikzpicture}
    \caption{Token cost in offline indexing using LLaMA3.0-8B (Clue-RAG-0.0 requires \textcolor{red}{no LLM}).}
    \label{fig:token_compare_optimized}
\end{figure}
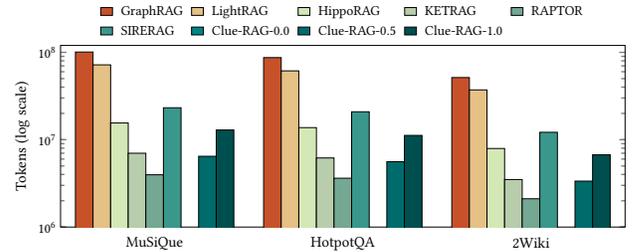

\subsection{Effectiveness of hybrid extraction (Q2)}
\label{exp:selection}

In this experiment, we evaluate the effectiveness of our hybrid extraction strategy for knowledge units, as described in Section~\ref{Clue-index}.
Table~\ref{tab:selection} reports the QA performance of four methods, each constrained to use only 50\% of the total token budget for LLM processing ($\alpha = 0.5$).
The methods differ in their chunk selection strategies: RANDOM (random selection), COS (our {\ttfamily Chunk-Selection} using cosine similarity), and BLEU (our {\ttfamily Chunk-Selection} using BLEU score).
For direct comparison, we also include the performance of Clue-RAG-1.0 (BLEU), which uses 100\% of the token budget.
The results show that both COS and BLEU significantly outperform RANDOM on average F1 and Acc.
Particularly, the BLEU-based strategy achieves the highest performance, exceeding RANDOM and COS by 3.14\% and 0.46\% in average F1, and by 3.05\% and 0.21\% in average Acc, respectively.
These gains underscore the effectiveness of our hybrid extraction approach in the offline indexing stage.

While Clue-RAG-1.0 (BLEU) yields stronger results, Clue-RAG-0.5 (BLEU) achieves 94.95\%–100.51\% of its performance, with half of the token usage.
An interesting phenomenon emerges on the MuSiQue dataset with LLaMA3.0-8B and Qwen2.5-32B, where Clue-RAG-0.5 (BLEU) occasionally outperforms its full-token counterpart.
This suggests that certain questions in MuSiQue exhibit a stronger lexical alignment with original sentences than with LLM-extracted knowledge units, as the paraphrasing process may substitute critical terms with less relevant alternatives.
Overall, these results demonstrate that hybrid extraction strategy can maintain robust QA performance while significantly reducing the token cost.






\definecolor{c0}{HTML}{642915}  
\definecolor{c1}{HTML}{963e20}  
\definecolor{c2}{HTML}{c7522a}  
\definecolor{c3}{HTML}{d68a58}  
\definecolor{c4}{HTML}{e5c185}  
\definecolor{c5}{HTML}{fbf2c4}  
\definecolor{c6}{HTML}{d4e8b7}  
\definecolor{c7}{HTML}{b8cdab}  
\definecolor{c8}{HTML}{74a892}  
\definecolor{c9}{HTML}{3a978c}  
\definecolor{c10}{HTML}{008585} 
\definecolor{c11}{HTML}{006b6b} 
\definecolor{c12}{HTML}{004f4f} 

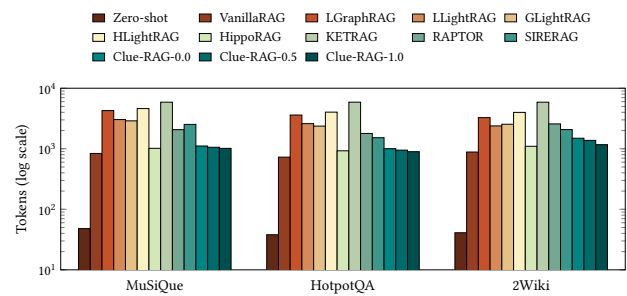
\begin{figure}[htbp]
    \centering
    \setlength{\abovecaptionskip}{2pt}
    \setlength{\belowcaptionskip}{-5pt}
    \begin{tikzpicture}[scale=0.52]
        \begin{axis}[
            ybar=0.0cm,
            bar width=0.3cm,
            width=0.9\textwidth,
            height=0.35\textwidth,
            xtick={1,2,3},
            xticklabels={MuSiQue, HotpotQA, 2Wiki},
            xticklabel style={font=\Large},
            yticklabel style={font=\large},
            legend style={
                at={(0.5,1.45)},
                anchor=north,
                legend columns=5,
                draw=none,
                font=\large,
                cells={anchor=west},
                column sep=4pt,
                row sep=2pt,
            },
            legend image code/.code={
                \draw [#1] (0cm,-0.1cm) rectangle (0.3cm,0.1cm);
            },
            xmin=0.5, xmax=3.5,
            ymin=10, ymax=10000,
            ymode=log,
            ytick={10,100,1000,10000},
            yticklabels={$10^1$,$10^2$,$10^3$,$10^4$},
            ylabel={\Large Tokens (log scale)},
            axis line style={line width=0.5pt},
            tick style={line width=0.5pt},
            xtick style={draw=none},
        ]
        
        \addplot[fill=c0, draw=black,line width=0.5pt] coordinates {(1,48) (2,38) (3,41)}; 
        \addplot[fill=c1, draw=black,line width=0.5pt] coordinates {(1,837) (2,728) (3,884)}; 
        \addplot[fill=c2, draw=black,line width=0.5pt] coordinates {(1,4270) (2,3607) (3,3270)}; 
        \addplot[fill=c3, draw=black,line width=0.5pt] coordinates {(1,3036) (2,2601) (3,2384)}; 
        \addplot[fill=c4, draw=black,line width=0.5pt] coordinates {(1,2892) (2,2367) (3,2542)}; 
        \addplot[fill=c5, draw=black,line width=0.5pt] coordinates {(1,4613) (2,4029) (3,3990)}; 
        \addplot[fill=c6, draw=black,line width=0.5pt] coordinates {(1,1017) (2,926) (3,1099)}; 
        \addplot[fill=c7, draw=black,line width=0.5pt] coordinates {(1,5880) (2,5881) (3,5879)}; 
        \addplot[fill=c8, draw=black,line width=0.5pt] coordinates {(1,2069) (2,1789) (3,2574)}; 
        \addplot[fill=c9, draw=black,line width=0.5pt] coordinates {(1,2528) (2,1523) (3,2075)}; 
        \addplot[fill=c10, draw=black,line width=0.5pt] coordinates {(1,1111) (2,1004) (3,1494)}; 
        \addplot[fill=c11, draw=black,line width=0.5pt] coordinates {(1,1059) (2,949) (3,1376)}; 
        \addplot[fill=c12, draw=black,line width=0.5pt] coordinates {(1,1016) (2,895) (3,1169)}; 

        \legend{
            Zero-shot,
            VanillaRAG,
            LGraphRAG,
            LLightRAG,
            GLightRAG,
            HLightRAG,
            HippoRAG,
            KETRAG,
            RAPTOR,
            SIRERAG,
            Clue-RAG-0.0,
            Clue-RAG-0.5,
            Clue-RAG-1.0
        }
        \end{axis}
    \end{tikzpicture}
    \caption{Token cost in online retrieval using LLaMA3.0-8B.}
    \label{fig:query_token}
\end{figure}

\subsection{Ablation study (Q3)}
\label{exp:ablation}

\begin{table*}[htbp]
\centering
\small
\caption{Ablation study of Clue-RAG-1.0, evaluating the contributions of three key components (K-A, S-A, and Q-U). The best performance for each metric is highlighted in bold, while the second-best result is underlined.}
\renewcommand{\arraystretch}{1.0}
\begin{tabular}{lcccccccccccccc}
\toprule
& \multicolumn{6}{c}{LLaMA3.0-8B} & \multicolumn{6}{c}{Qwen2.5-32B} & \multicolumn{2}{c}{}\\
\cmidrule(lr){2-7} \cmidrule(lr){8-13} 
 & \multicolumn{2}{c}{MuSiQue} & \multicolumn{2}{c}{HotpotQA} & \multicolumn{2}{c}{2Wiki} & \multicolumn{2}{c}{MuSiQue} & \multicolumn{2}{c}{HotpotQA} & \multicolumn{2}{c}{2Wiki}  & \multicolumn{2}{c}{Avg}\\ 
\cmidrule(lr){2-3} \cmidrule(lr){4-5} \cmidrule(lr){6-7} \cmidrule(lr){8-9} \cmidrule(lr){10-11} \cmidrule(lr){12-13} \cmidrule(lr){14-15}
Method & F1 & Acc. & F1 & Acc. & F1 & Acc. & F1 & Acc. & F1 & Acc. & F1 & Acc. & F1 & Acc. \\ 
\midrule

{\tt Clue-RAG-1.0} & \textbf{25.48} & \textbf{22.32} & \textbf{55.97} & \textbf{59.20} & \underline{37.41} & \underline{50.70} & \textbf{36.50} & \textbf{32.83} & \textbf{64.02} & \textbf{64.70} & \underline{55.20} & \underline{61.00} & \textbf{45.76} & \textbf{48.46} \\

{\tt w/o K-A} & \underline{25.26} & \underline{22.02} & 52.93 & 55.50 & 36.75 & 50.10 & 32.64 & 29.13 & 60.50 & 62.00 & 52.58 & 58.60 & 43.44 & 46.23 \\

{\tt w/o S-A} & 23.18 & 19.52 & 53.48 & 55.70 & 34.76 & 45.60 & 31.64 & 28.03 & 59.12 & 59.90 & 48.57 & 55.70 & 41.79 & 44.08 \\

{\tt w/o Q-U} & 24.51 & 20.72 & \underline{54.51} & \underline{56.80} & \textbf{38.00} & \textbf{51.50} & \underline{35.36} & \underline{32.03} & \underline{62.52} & \underline{64.10} & \textbf{55.27} & \textbf{61.80} & \underline{45.03} & \underline{47.83} \\

{\tt \makecell[l]{w/o Q-U or K-A}} & 22.53 & 18.82 & 52.00 & 53.90 & 36.52 & 49.50 & 31.93 & 28.33 & 60.14 & 62.00 & 52.72 & 58.90 & 42.64 & 45.24 \\

{\tt \makecell[l]{w/o Q-U or S-A}} & 21.57 & 17.42 & 51.22 & 53.50 & 33.43 & 46.10 & 31.84 & 27.93 & 56.71 & 58.30 & 48.34 & 55.80 & 40.52 & 43.18 \\

\bottomrule
\end{tabular}
\label{tab:ablation}
\end{table*}

In this experiment, we evaluate the key components of {\tt Q-Iter}, i.e., Spreading Activation (S-A below) and Knowledge Anchoring (K-A below) of Step-1, and the Query Updating (Q-U below) of Step-2.
Table \ref{tab:ablation} shows the performance of removing each component.
We observe that S-A enhances the average F1 and Acc. by 9.5\% and 9.95\%, respectively, while K-A provides improvement of 5.34\% and 4.83\%, and Q-U improves the average F1 and Acc. by 1.63\% and 1.32\%.
The superior performance of S-A can be attributed to its alignment in semantic granularity: The entities extracted from the query by the LLM are matched against entity nodes in the multi-partite graph, ensuring consistency.
In contrast, K-A relies on semantic embeddings of the query to align with knowledge unit nodes in the graph, which may introduce noise due to potential inconsistencies in semantic granularity.
Q-U improves F1 by up to 3.96\% and Acc by 7.72\% on MuSiQue, and by 2.68\% and 4.23\% on HotpotQA, but slightly degrades on 2Wiki. The performance decline on 2Wiki stems from its frequent semantically similar entities (e.g., family relations or homonyms like Elizabeth I/II), which challenge disambiguation.
Nevertheless, they collectively enhance the retrieval performance of Clue-RAG.


\subsection{Parameter sensitivity (Q4)}
In this experiment, we analyze the sensitivity of Clue-RAG using Qwen2.5-32B w.r.t. key hyperparameters: the number $K$ of top-$K$ retrieved results, beam size $M$, search depth $D$, and token constraint coefficient $\alpha$, where $K \in \{3,5,7,9\}$, $M \in \{3,5,7,9\}$, $D \in \{1,2,3,4\}$, and $\alpha \in \{0,0.25,0.5,0.75,1.0\}$.

First, as shown in Figure \ref{fig:hyperparameter} (a), Clue-RAG does not exhibit a consistent improvement in generation quality with increasing $K$ across all three datasets. Instead, the results suggest that $K=3$ provides sufficient retrieval results, while larger values introduce noise without meaningful gains.
Next, Figure \ref{fig:hyperparameter} (b) reveals that generation quality generally improves with larger beam sizes $M$. However, the gains plateau beyond $M=5$, and a slight decline occurs at $M=7,9$. This indicates that $M=5$ strikes an optimal balance between generation quality and computational efficiency, as excessively large beam sizes may introduce noise with diminishing returns.
Furthermore, Figure \ref{fig:hyperparameter} (c) demonstrates that deeper search depth $D$ leads to better performance, which aligns with the nature of the datasets, where most questions require 2 hops while fewer necessitate 3-4 hops. Nevertheless, when $D=4$, generation quality slightly deteriorates, likely due to over-retrieval of irrelevant information for most 2-hops questions.
Finally, Figure \ref{fig:hyperparameter} (d) shows a positive correlation between the token constraint coefficient $\alpha$ and generation quality. This is expected because higher $\alpha$ values allow the LLM to process more chunks, thereby improving the quality of knowledge units and enhancing retrieval accuracy. 
Consistent results are observed across other datasets in Appendix Figure \ref{fig:hyperparameter_appendix}.

\definecolor{ch1}{HTML}{c7522a} 
\definecolor{ch2}{HTML}{008585} 
\begin{figure}[htbp] 
    \centering
\begin{tikzpicture}
    \begin{groupplot}[
        group style={
            group size=4 by 1,
            horizontal sep=0.5cm,
        },
        width=0.18\textwidth,
        height=0.15\textwidth,
        tick label style={font=\scriptsize},           
    ]

    \nextgroupplot[
        title={}, ymin=31, ymax=37, 
        ylabel={},
        ytick={31, 34, 37},
        ylabel near ticks,
        title style={yshift=-5pt},
        xmin=3, xmax=9,
        xtick={3,5,7,9},
        xlabel={\textbf{(a) Top-$\mathbf{\textit{K}}$}}
    ]
    \addplot[
        ch1, 
        mark=*, 
        mark options={solid, fill=white},  
        line width=1.0pt,                 
    ] coordinates {(3,36.50)(5,36.14)(7,36.23)(9,36.45)};
    \addplot[
        ch2, 
        mark=triangle*, 
        mark options={solid, fill=white},  
        line width=1.0pt,                 
    ] coordinates {(3,32.83)(5,32.53)(7,32.63)(9,33.03)};


    \nextgroupplot[ymin=31, ymax=37, ytick={31, 34, 37},ylabel near ticks,xmin=3, xmax=9,
    xtick={3,5,7,9},ylabel={},xlabel={\textbf{(b) $\mathbf{\textit{M}}$}}, legend style={at={(0.5,-0.7)}, anchor=north, legend columns=-1}]

    \addplot[
            ch1, 
            mark=*, 
            mark options={solid, fill=white},  
            line width=1.0pt,                 
        ] coordinates {(3,34.79)(5,36.50)(7,35.75)(9,35.28)};

\addlegendimage{ch1, mark=triangle*, mark options={solid, fill=white}, line width=1.0pt}

    \addplot[
            ch2, 
            mark=triangle*, 
            mark options={solid, fill=white},  
            line width=1.0pt,                 
        ] coordinates {(3,31.13)(5,32.83)(7,32.23)(9,31.73)};

    \nextgroupplot[title={MuSiQue},ymin=20, ymax=40, ytick={20, 30, 40}, ylabel near ticks,xmin=1, xmax=4,
    xtick={1,2,3,4},ylabel={},xlabel={\textbf{(c) $\mathbf{\textit{D}}$}}, legend style={at={(-0.1,-0.85)}, anchor=north, legend columns=-1},title style={yshift=-1mm,xshift=-10mm,}]

\addplot[
        ch1, 
        mark=*, 
        mark options={solid, fill=white},  
        line width=1.0pt,                 
    ] coordinates {(1,26.45)(2,34.73)(3,36.50)(4,36.15)};

\addlegendentry{F1}

\addplot[
        ch2, 
        mark=triangle*, 
        mark options={solid, fill=white},  
        line width=1.0pt,                 
    ] coordinates {(1,22.12)(2,31.03)(3,32.83)(4,32.73)};

\addlegendentry{Acc.}

\nextgroupplot[ymin=30, ymax=38, ytick={30, 34, 38}, ylabel near ticks,xmin=0.0, xmax=1.0,
    xtick={0.0, 0.25, 0.5, 0.75, 1.0},ylabel={},xlabel={\textbf{(d) $\mathbf{\alpha}$}}]

\addplot[
    ch1, 
    mark=*, 
    mark options={solid, fill=white},  
    line width=1.0pt                  
] coordinates {
    (0.0,34.26)
    (0.25,36.21)
    (0.5,36.68)
    (0.75,36.55)
    (1.0,36.50)
};

\addplot[
    ch2,
    mark=triangle*,
    mark options={solid, fill=white},  
    line width=1.0pt                  
] coordinates {
    (0.0,31.33)
    (0.25,32.63)
    (0.5,32.93)
    (0.75,34.03)
    (1.0,32.83)
};

    \end{groupplot}
\end{tikzpicture}
\vspace*{-20pt}
\caption{Performance analysis under different hyperparameter settings on MuSiQue dataset.}
\label{fig:hyperparameter}
\end{figure}
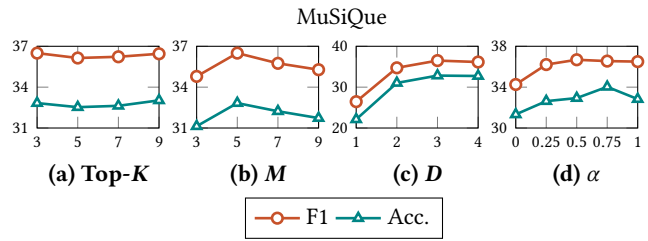

\section{Related Works}
\label{sec:related}

In this section, we review the representative RAG methods based on vector database (VDB) \cite{pan2024survey, zhang2024there}, trees, and graphs.
For more details, please see recent surveys \cite{peng2024graph, zhang2025survey, fan2024survey} and empirical study \cite{zhou2025depth}. 

$\bullet$ {\bf VDB-based RAG.}
Vallina RAG is a basic RAG solution that (1) splits the corpora into text chunks, (2) encodes them into embeddings via an embedding model, and (3) stores embeddings in a VDB.
During online retrieval, the same model embeds the query and then retrieves top-$K$ relevant text chunks for augmented generation.

$\bullet$ {\bf Tree-based RAG.}
RAPTOR \cite{sarthi2024raptor} introduces a tree-based index, which can be built in a bottom-up tree manner. Specifically, it begins with raw text chunks as leaf nodes, clusters their embeddings via Gaussian Mixture Models \cite{mcinnes2018umap}, then recursively summarizes clusters using LLMs and re-clusters them to build the tree bottom-up.
During online retrieval, all tree nodes are indexed in a VDB, enabling a fast semantic similarity search with query embeddings.
Building on RAPTOR, SIRERAG \cite{zhang2024sirerag} introduces a dual tree framework, which first extracts propositions and entities from texts using LLMs, regrouping propositions linked to the same entity into new passages, and then organize these passages into a relatedness tree alongside the RAPTOR similarity tree, with both trees jointly indexed in the same vector database for retrieval.
This hybrid approach aims to balance semantic similarity and relatedness coherence. 
EraRAG \cite{zhang2025eraragefficientincrementalretrieval} is a novel hierarchical tree-based RAG method that focuses on the dynamic scenario, allowing for efficient and incremental index updates as the corpus evolves.

$\bullet$ {\bf Graph-based RAG.}
Compared to trees, graphs are more effective for modeling complicated relationships.
Microsoft’s GraphRAG \cite{edge2024local} first constructs a text-attributed KG by extracting entities, relationships, and other detailed contextual features from text fragments through LLMs.
Then, it employs Leiden algorithm \cite{traag2019louvain} to cluster the KG into some hierarchical clusters, where each cluster is associated with a community report.
For retrieval, it uses the local search to retrieve relevant context from entities, relationships, community reports, and text chunks to augment response.
ArchRAG \cite{wang2025archragattributedcommunitybasedhierarchical} organizes attributed communities hierarchically, and augments the question using summaries of attributed communities.
LightRAG \cite{guo2024lightrag} constructs a KG from text chunks while using LLMs to augment both entities and relations with extracted keywords and leverages LLMs to extract query-relevant low-/high-level keywords for retrieval.
%
%
KETRAG \cite{huang2025ket} first selects core text chunks from a KNN graph built using both semantic and lexical similarity, and then constructs a KG. It supports online retrieval by using the standard local search but extracting ego networks to improve LLM generation.
HippoRAG \cite{gutierrez2024hipporag} constructs a KG by extracting triplets from the text chunks using LLMs and enhances its quality by linking similar entities via semantic embeddings. For multi-hop reasoning, it employs PPR to retrieve relevant KG entities and augments the response with entities' associated text chunks for LLM generation.

\section{Conclusion}
\label{sec:conclude}

In this work, we propose Clue-RAG, a graph-based RAG approach that features a multi-partite graph index integrating chunks, knowledge units, and entities.
To efficiently build this index, we introduce a hybrid extraction strategy for knowledge unit that maximizes LLM processing benefits while minimizing token usage.
For online retrieval, we design a query-driven iterative retrieval ({\ttfamily Q-Iter}) that ensures relevant retrieval results.
These designs enhance graph quality and retrieval relevance while reducing LLM token consumption.
Experimental results on three QA benchmarks demonstrate that Clue-RAG significantly outperforms SOTA baselines in both QA performance and cost efficiency.
Notably, its zero-token variant achieves comparable or superior performance, highlighting the robustness and effectiveness of our approach.
In the future, we will extend our Clue-RAG for processing multimodal data.

\clearpage

\balance
\bibliographystyle{ACM-Reference-Format}
\bibliography{sample-base}


\begin{thebibliography}{49}


\ifx \showCODEN    \undefined \def \showCODEN     #1{\unskip}     \fi
\ifx \showISBNx    \undefined \def \showISBNx     #1{\unskip}     \fi
\ifx \showISBNxiii \undefined \def \showISBNxiii  #1{\unskip}     \fi
\ifx \showISSN     \undefined \def \showISSN      #1{\unskip}     \fi
\ifx \showLCCN     \undefined \def \showLCCN      #1{\unskip}     \fi
\ifx \shownote     \undefined \def \shownote      #1{#1}          \fi
\ifx \showarticletitle \undefined \def \showarticletitle #1{#1}   \fi
\ifx \showURL      \undefined \def \showURL       {\relax}        \fi
\providecommand\bibfield[2]{#2}
\providecommand\bibinfo[2]{#2}
\providecommand\natexlab[1]{#1}
\providecommand\showeprint[2][]{arXiv:#2}

\bibitem[Anderson(1983)]%
        {anderson1983spreading}
\bibfield{author}{\bibinfo{person}{John~R Anderson}.} \bibinfo{year}{1983}\natexlab{}.
\newblock \showarticletitle{A spreading activation theory of memory}.
\newblock \bibinfo{journal}{\emph{Journal of verbal learning and verbal behavior}} \bibinfo{volume}{22}, \bibinfo{number}{3} (\bibinfo{year}{1983}), \bibinfo{pages}{261--295}.
\newblock


\bibitem[Asai et~al\mbox{.}(2023)]%
        {asai2023self}
\bibfield{author}{\bibinfo{person}{Akari Asai}, \bibinfo{person}{Zeqiu Wu}, \bibinfo{person}{Yizhong Wang}, \bibinfo{person}{Avirup Sil}, {and} \bibinfo{person}{Hannaneh Hajishirzi}.} \bibinfo{year}{2023}\natexlab{}.
\newblock \showarticletitle{Self-rag: Learning to retrieve, generate, and critique through self-reflection}. In \bibinfo{booktitle}{\emph{The Twelfth International Conference on Learning Representations}}.
\newblock


\bibitem[Bai et~al\mbox{.}(2023)]%
        {bai2023qwen}
\bibfield{author}{\bibinfo{person}{Jinze Bai}, \bibinfo{person}{Shuai Bai}, \bibinfo{person}{Yunfei Chu}, \bibinfo{person}{Zeyu Cui}, \bibinfo{person}{Kai Dang}, \bibinfo{person}{Xiaodong Deng}, \bibinfo{person}{Yang Fan}, \bibinfo{person}{Wenbin Ge}, \bibinfo{person}{Yu Han}, \bibinfo{person}{Fei Huang}, {et~al\mbox{.}}} \bibinfo{year}{2023}\natexlab{}.
\newblock \showarticletitle{Qwen technical report}.
\newblock \bibinfo{journal}{\emph{arXiv preprint arXiv:2309.16609}} (\bibinfo{year}{2023}).
\newblock


\bibitem[Bird et~al\mbox{.}(2009)]%
        {Bird_Natural_Language_Processing_2009}
\bibfield{author}{\bibinfo{person}{Steven Bird}, \bibinfo{person}{Ewan Klein}, {and} \bibinfo{person}{Edward Loper}.} \bibinfo{year}{2009}\natexlab{}.
\newblock \bibinfo{booktitle}{\emph{{Natural Language Processing with Python: Analyzing Text with the Natural Language Toolkit}}}.
\newblock \bibinfo{publisher}{O'Reilly Media, Inc.}
\newblock
\showISBNx{9780596516499}
\urldef\tempurl%
\url{https://www.nltk.org/book/}
\showURL{%
\tempurl}


\bibitem[Chen et~al\mbox{.}(2024)]%
        {chen2024bge}
\bibfield{author}{\bibinfo{person}{Jianlv Chen}, \bibinfo{person}{Shitao Xiao}, \bibinfo{person}{Peitian Zhang}, \bibinfo{person}{Kun Luo}, \bibinfo{person}{Defu Lian}, {and} \bibinfo{person}{Zheng Liu}.} \bibinfo{year}{2024}\natexlab{}.
\newblock \bibinfo{title}{BGE M3-Embedding: Multi-Lingual, Multi-Functionality, Multi-Granularity Text Embeddings Through Self-Knowledge Distillation}.
\newblock
\showeprint[arxiv]{2402.03216}~[cs.CL]


\bibitem[Collins and Loftus(1975)]%
        {collins1975spreading}
\bibfield{author}{\bibinfo{person}{Allan~M Collins} {and} \bibinfo{person}{Elizabeth~F Loftus}.} \bibinfo{year}{1975}\natexlab{}.
\newblock \showarticletitle{A spreading-activation theory of semantic processing.}
\newblock \bibinfo{journal}{\emph{Psychological review}} \bibinfo{volume}{82}, \bibinfo{number}{6} (\bibinfo{year}{1975}), \bibinfo{pages}{407}.
\newblock


\bibitem[DeepSeek-AI(2025)]%
        {deepseekai2025deepseekv3technicalreport}
\bibfield{author}{\bibinfo{person}{DeepSeek-AI}.} \bibinfo{year}{2025}\natexlab{}.
\newblock \bibinfo{title}{DeepSeek-V3 Technical Report}.
\newblock
\showeprint[arxiv]{2412.19437}~[cs.CL]
\urldef\tempurl%
\url{https://arxiv.org/abs/2412.19437}
\showURL{%
\tempurl}


\bibitem[Edge et~al\mbox{.}(2024)]%
        {edge2024local}
\bibfield{author}{\bibinfo{person}{Darren Edge}, \bibinfo{person}{Ha Trinh}, \bibinfo{person}{Newman Cheng}, \bibinfo{person}{Joshua Bradley}, \bibinfo{person}{Alex Chao}, \bibinfo{person}{Apurva Mody}, \bibinfo{person}{Steven Truitt}, \bibinfo{person}{Dasha Metropolitansky}, \bibinfo{person}{Robert~Osazuwa Ness}, {and} \bibinfo{person}{Jonathan Larson}.} \bibinfo{year}{2024}\natexlab{}.
\newblock \showarticletitle{From local to global: A graph rag approach to query-focused summarization}.
\newblock \bibinfo{journal}{\emph{arXiv preprint arXiv:2404.16130}} (\bibinfo{year}{2024}).
\newblock


\bibitem[Fan et~al\mbox{.}(2024)]%
        {fan2024survey}
\bibfield{author}{\bibinfo{person}{Wenqi Fan}, \bibinfo{person}{Yujuan Ding}, \bibinfo{person}{Liangbo Ning}, \bibinfo{person}{Shijie Wang}, \bibinfo{person}{Hengyun Li}, \bibinfo{person}{Dawei Yin}, \bibinfo{person}{Tat-Seng Chua}, {and} \bibinfo{person}{Qing Li}.} \bibinfo{year}{2024}\natexlab{}.
\newblock \showarticletitle{A survey on rag meeting llms: Towards retrieval-augmented large language models}. In \bibinfo{booktitle}{\emph{Proceedings of the 30th ACM SIGKDD Conference on Knowledge Discovery and Data Mining}}. \bibinfo{pages}{6491--6501}.
\newblock


\bibitem[Gao et~al\mbox{.}(2023)]%
        {gao2023retrieval}
\bibfield{author}{\bibinfo{person}{Yunfan Gao}, \bibinfo{person}{Yun Xiong}, \bibinfo{person}{Xinyu Gao}, \bibinfo{person}{Kangxiang Jia}, \bibinfo{person}{Jinliu Pan}, \bibinfo{person}{Yuxi Bi}, \bibinfo{person}{Yixin Dai}, \bibinfo{person}{Jiawei Sun}, \bibinfo{person}{Haofen Wang}, {and} \bibinfo{person}{Haofen Wang}.} \bibinfo{year}{2023}\natexlab{}.
\newblock \showarticletitle{Retrieval-augmented generation for large language models: A survey}.
\newblock \bibinfo{journal}{\emph{arXiv preprint arXiv:2312.10997}} \bibinfo{volume}{2}, \bibinfo{number}{1} (\bibinfo{year}{2023}).
\newblock


\bibitem[Ghimire et~al\mbox{.}(2024)]%
        {ghimire2024generative}
\bibfield{author}{\bibinfo{person}{Aashish Ghimire}, \bibinfo{person}{James Pather}, {and} \bibinfo{person}{John Edwards}.} \bibinfo{year}{2024}\natexlab{}.
\newblock \showarticletitle{Generative AI in education: A study of Educators' awareness, sentiments, and influencing factors}. In \bibinfo{booktitle}{\emph{2024 IEEE Frontiers in Education Conference (FIE)}}. IEEE, \bibinfo{pages}{1--9}.
\newblock


\bibitem[Grattafiori et~al\mbox{.}(2024)]%
        {grattafiori2024llama}
\bibfield{author}{\bibinfo{person}{Aaron Grattafiori}, \bibinfo{person}{Abhimanyu Dubey}, \bibinfo{person}{Abhinav Jauhri}, \bibinfo{person}{Abhinav Pandey}, \bibinfo{person}{Abhishek Kadian}, \bibinfo{person}{Ahmad Al-Dahle}, \bibinfo{person}{Aiesha Letman}, \bibinfo{person}{Akhil Mathur}, \bibinfo{person}{Alan Schelten}, \bibinfo{person}{Alex Vaughan}, {et~al\mbox{.}}} \bibinfo{year}{2024}\natexlab{}.
\newblock \showarticletitle{The llama 3 herd of models}.
\newblock \bibinfo{journal}{\emph{arXiv preprint arXiv:2407.21783}} (\bibinfo{year}{2024}).
\newblock


\bibitem[Guo et~al\mbox{.}(2024)]%
        {guo2024lightrag}
\bibfield{author}{\bibinfo{person}{Zirui Guo}, \bibinfo{person}{Lianghao Xia}, \bibinfo{person}{Yanhua Yu}, \bibinfo{person}{Tu Ao}, {and} \bibinfo{person}{Chao Huang}.} \bibinfo{year}{2024}\natexlab{}.
\newblock \showarticletitle{Lightrag: Simple and fast retrieval-augmented generation}.
\newblock  (\bibinfo{year}{2024}).
\newblock


\bibitem[Guti{\'e}rrez et~al\mbox{.}(2024)]%
        {gutierrez2024hipporag}
\bibfield{author}{\bibinfo{person}{Bernal~Jim{\'e}nez Guti{\'e}rrez}, \bibinfo{person}{Yiheng Shu}, \bibinfo{person}{Yu Gu}, \bibinfo{person}{Michihiro Yasunaga}, {and} \bibinfo{person}{Yu Su}.} \bibinfo{year}{2024}\natexlab{}.
\newblock \showarticletitle{Hipporag: Neurobiologically inspired long-term memory for large language models}. In \bibinfo{booktitle}{\emph{The Thirty-eighth Annual Conference on Neural Information Processing Systems}}.
\newblock


\bibitem[Ho et~al\mbox{.}(2020)]%
        {ho2020constructing}
\bibfield{author}{\bibinfo{person}{Xanh Ho}, \bibinfo{person}{Anh-Khoa~Duong Nguyen}, \bibinfo{person}{Saku Sugawara}, {and} \bibinfo{person}{Akiko Aizawa}.} \bibinfo{year}{2020}\natexlab{}.
\newblock \showarticletitle{Constructing a multi-hop qa dataset for comprehensive evaluation of reasoning steps}.
\newblock \bibinfo{journal}{\emph{arXiv preprint arXiv:2011.01060}} (\bibinfo{year}{2020}).
\newblock


\bibitem[Honnibal et~al\mbox{.}(2020)]%
        {Honnibal_spaCy_Industrial-strength_Natural_2020}
\bibfield{author}{\bibinfo{person}{Matthew Honnibal}, \bibinfo{person}{Ines Montani}, \bibinfo{person}{Sofie Van~Landeghem}, {and} \bibinfo{person}{Adriane Boyd}.} \bibinfo{year}{2020}\natexlab{}.
\newblock \showarticletitle{{spaCy: Industrial-strength Natural Language Processing in Python}}.
\newblock  (\bibinfo{year}{2020}).
\newblock
\href{https://doi.org/10.5281/zenodo.1212303}{doi:\nolinkurl{10.5281/zenodo.1212303}}


\bibitem[Hu and Lu(2024)]%
        {hu2024rag}
\bibfield{author}{\bibinfo{person}{Yucheng Hu} {and} \bibinfo{person}{Yuxing Lu}.} \bibinfo{year}{2024}\natexlab{}.
\newblock \showarticletitle{Rag and rau: A survey on retrieval-augmented language model in natural language processing}.
\newblock \bibinfo{journal}{\emph{arXiv preprint arXiv:2404.19543}} (\bibinfo{year}{2024}).
\newblock


\bibitem[Huang and Huang(2024)]%
        {huang2024survey}
\bibfield{author}{\bibinfo{person}{Yizheng Huang} {and} \bibinfo{person}{Jimmy Huang}.} \bibinfo{year}{2024}\natexlab{}.
\newblock \showarticletitle{A survey on retrieval-augmented text generation for large language models}.
\newblock \bibinfo{journal}{\emph{arXiv preprint arXiv:2404.10981}} (\bibinfo{year}{2024}).
\newblock


\bibitem[Huang et~al\mbox{.}(2025)]%
        {huang2025ket}
\bibfield{author}{\bibinfo{person}{Yiqian Huang}, \bibinfo{person}{Shiqi Zhang}, {and} \bibinfo{person}{Xiaokui Xiao}.} \bibinfo{year}{2025}\natexlab{}.
\newblock \showarticletitle{KET-RAG: A Cost-Efficient Multi-Granular Indexing Framework for Graph-RAG}.
\newblock \bibinfo{journal}{\emph{arXiv preprint arXiv:2502.09304}} (\bibinfo{year}{2025}).
\newblock


\bibitem[Liu et~al\mbox{.}(2024)]%
        {liu2024survey}
\bibfield{author}{\bibinfo{person}{Lei Liu}, \bibinfo{person}{Xiaoyan Yang}, \bibinfo{person}{Junchi Lei}, \bibinfo{person}{Xiaoyang Liu}, \bibinfo{person}{Yue Shen}, \bibinfo{person}{Zhiqiang Zhang}, \bibinfo{person}{Peng Wei}, \bibinfo{person}{Jinjie Gu}, \bibinfo{person}{Zhixuan Chu}, \bibinfo{person}{Zhan Qin}, {et~al\mbox{.}}} \bibinfo{year}{2024}\natexlab{}.
\newblock \showarticletitle{A survey on medical large language models: Technology, application, trustworthiness, and future directions}.
\newblock \bibinfo{journal}{\emph{arXiv preprint arXiv:2406.03712}} (\bibinfo{year}{2024}).
\newblock


\bibitem[Liu et~al\mbox{.}(2023)]%
        {liu2023lost}
\bibfield{author}{\bibinfo{person}{Nelson~F Liu}, \bibinfo{person}{Kevin Lin}, \bibinfo{person}{John Hewitt}, \bibinfo{person}{Ashwin Paranjape}, \bibinfo{person}{Michele Bevilacqua}, \bibinfo{person}{Fabio Petroni}, {and} \bibinfo{person}{Percy Liang}.} \bibinfo{year}{2023}\natexlab{}.
\newblock \showarticletitle{Lost in the middle: How language models use long contexts}.
\newblock \bibinfo{journal}{\emph{arXiv preprint arXiv:2307.03172}} (\bibinfo{year}{2023}).
\newblock


\bibitem[Mallen et~al\mbox{.}(2022)]%
        {mallen2022not}
\bibfield{author}{\bibinfo{person}{Alex Mallen}, \bibinfo{person}{Akari Asai}, \bibinfo{person}{Victor Zhong}, \bibinfo{person}{Rajarshi Das}, \bibinfo{person}{Daniel Khashabi}, {and} \bibinfo{person}{Hannaneh Hajishirzi}.} \bibinfo{year}{2022}\natexlab{}.
\newblock \showarticletitle{When not to trust language models: Investigating effectiveness of parametric and non-parametric memories}.
\newblock \bibinfo{journal}{\emph{arXiv preprint arXiv:2212.10511}} (\bibinfo{year}{2022}).
\newblock


\bibitem[McInnes et~al\mbox{.}(2018)]%
        {mcinnes2018umap}
\bibfield{author}{\bibinfo{person}{Leland McInnes}, \bibinfo{person}{John Healy}, {and} \bibinfo{person}{James Melville}.} \bibinfo{year}{2018}\natexlab{}.
\newblock \showarticletitle{Umap: Uniform manifold approximation and projection for dimension reduction}.
\newblock \bibinfo{journal}{\emph{arXiv preprint arXiv:1802.03426}} (\bibinfo{year}{2018}).
\newblock


\bibitem[Nie et~al\mbox{.}(2024)]%
        {nie2024survey}
\bibfield{author}{\bibinfo{person}{Yuqi Nie}, \bibinfo{person}{Yaxuan Kong}, \bibinfo{person}{Xiaowen Dong}, \bibinfo{person}{John~M Mulvey}, \bibinfo{person}{H~Vincent Poor}, \bibinfo{person}{Qingsong Wen}, {and} \bibinfo{person}{Stefan Zohren}.} \bibinfo{year}{2024}\natexlab{}.
\newblock \showarticletitle{A survey of large language models for financial applications: Progress, prospects and challenges}.
\newblock \bibinfo{journal}{\emph{arXiv preprint arXiv:2406.11903}} (\bibinfo{year}{2024}).
\newblock


\bibitem[OpenAI({[n.\,d.]})]%
        {openai_tik}
\bibfield{author}{\bibinfo{person}{OpenAI}.} \bibinfo{year}{[n.\,d.]}\natexlab{}.
\newblock \bibinfo{title}{{tiktoken}}.
\newblock
\urldef\tempurl%
\url{https://github.com/openai/tiktoken}
\showURL{%
\tempurl}


\bibitem[Pan et~al\mbox{.}(2024)]%
        {pan2024survey}
\bibfield{author}{\bibinfo{person}{James~Jie Pan}, \bibinfo{person}{Jianguo Wang}, {and} \bibinfo{person}{Guoliang Li}.} \bibinfo{year}{2024}\natexlab{}.
\newblock \showarticletitle{Survey of vector database management systems}.
\newblock \bibinfo{journal}{\emph{The VLDB Journal}} \bibinfo{volume}{33}, \bibinfo{number}{5} (\bibinfo{year}{2024}), \bibinfo{pages}{1591--1615}.
\newblock


\bibitem[Papineni et~al\mbox{.}(2002)]%
        {papineni2002bleu}
\bibfield{author}{\bibinfo{person}{Kishore Papineni}, \bibinfo{person}{Salim Roukos}, \bibinfo{person}{Todd Ward}, {and} \bibinfo{person}{Wei-Jing Zhu}.} \bibinfo{year}{2002}\natexlab{}.
\newblock \showarticletitle{Bleu: a method for automatic evaluation of machine translation}. In \bibinfo{booktitle}{\emph{Proceedings of the 40th annual meeting of the Association for Computational Linguistics}}. \bibinfo{pages}{311--318}.
\newblock


\bibitem[Peng et~al\mbox{.}(2024)]%
        {peng2024graph}
\bibfield{author}{\bibinfo{person}{Boci Peng}, \bibinfo{person}{Yun Zhu}, \bibinfo{person}{Yongchao Liu}, \bibinfo{person}{Xiaohe Bo}, \bibinfo{person}{Haizhou Shi}, \bibinfo{person}{Chuntao Hong}, \bibinfo{person}{Yan Zhang}, {and} \bibinfo{person}{Siliang Tang}.} \bibinfo{year}{2024}\natexlab{}.
\newblock \showarticletitle{Graph retrieval-augmented generation: A survey}.
\newblock \bibinfo{journal}{\emph{arXiv preprint arXiv:2408.08921}} (\bibinfo{year}{2024}).
\newblock


\bibitem[Press et~al\mbox{.}(2022)]%
        {press2022measuring}
\bibfield{author}{\bibinfo{person}{Ofir Press}, \bibinfo{person}{Muru Zhang}, \bibinfo{person}{Sewon Min}, \bibinfo{person}{Ludwig Schmidt}, \bibinfo{person}{Noah~A Smith}, {and} \bibinfo{person}{Mike Lewis}.} \bibinfo{year}{2022}\natexlab{}.
\newblock \showarticletitle{Measuring and narrowing the compositionality gap in language models}.
\newblock \bibinfo{journal}{\emph{arXiv preprint arXiv:2210.03350}} (\bibinfo{year}{2022}).
\newblock


\bibitem[Ram et~al\mbox{.}(2023)]%
        {ram2023context}
\bibfield{author}{\bibinfo{person}{Ori Ram}, \bibinfo{person}{Yoav Levine}, \bibinfo{person}{Itay Dalmedigos}, \bibinfo{person}{Dor Muhlgay}, \bibinfo{person}{Amnon Shashua}, \bibinfo{person}{Kevin Leyton-Brown}, {and} \bibinfo{person}{Yoav Shoham}.} \bibinfo{year}{2023}\natexlab{}.
\newblock \showarticletitle{In-context retrieval-augmented language models}.
\newblock \bibinfo{journal}{\emph{Transactions of the Association for Computational Linguistics}}  \bibinfo{volume}{11} (\bibinfo{year}{2023}), \bibinfo{pages}{1316--1331}.
\newblock


\bibitem[Sahni(1975)]%
        {sahni1975approximate}
\bibfield{author}{\bibinfo{person}{Sartaj Sahni}.} \bibinfo{year}{1975}\natexlab{}.
\newblock \showarticletitle{Approximate algorithms for the 0/1 knapsack problem}.
\newblock \bibinfo{journal}{\emph{Journal of the ACM (JACM)}} \bibinfo{volume}{22}, \bibinfo{number}{1} (\bibinfo{year}{1975}), \bibinfo{pages}{115--124}.
\newblock


\bibitem[Sarthi et~al\mbox{.}(2024)]%
        {sarthi2024raptor}
\bibfield{author}{\bibinfo{person}{Parth Sarthi}, \bibinfo{person}{Salman Abdullah}, \bibinfo{person}{Aditi Tuli}, \bibinfo{person}{Shubh Khanna}, \bibinfo{person}{Anna Goldie}, {and} \bibinfo{person}{Christopher~D Manning}.} \bibinfo{year}{2024}\natexlab{}.
\newblock \showarticletitle{Raptor: Recursive abstractive processing for tree-organized retrieval}. In \bibinfo{booktitle}{\emph{The Twelfth International Conference on Learning Representations}}.
\newblock


\bibitem[Traag et~al\mbox{.}(2019)]%
        {traag2019louvain}
\bibfield{author}{\bibinfo{person}{Vincent~A Traag}, \bibinfo{person}{Ludo Waltman}, {and} \bibinfo{person}{Nees~Jan Van~Eck}.} \bibinfo{year}{2019}\natexlab{}.
\newblock \showarticletitle{From Louvain to Leiden: guaranteeing well-connected communities}.
\newblock \bibinfo{journal}{\emph{Scientific reports}} \bibinfo{volume}{9}, \bibinfo{number}{1} (\bibinfo{year}{2019}), \bibinfo{pages}{1--12}.
\newblock


\bibitem[Trivedi et~al\mbox{.}(2022a)]%
        {trivedi2022interleaving}
\bibfield{author}{\bibinfo{person}{Harsh Trivedi}, \bibinfo{person}{Niranjan Balasubramanian}, \bibinfo{person}{Tushar Khot}, {and} \bibinfo{person}{Ashish Sabharwal}.} \bibinfo{year}{2022}\natexlab{a}.
\newblock \showarticletitle{Interleaving retrieval with chain-of-thought reasoning for knowledge-intensive multi-step questions}.
\newblock \bibinfo{journal}{\emph{arXiv preprint arXiv:2212.10509}} (\bibinfo{year}{2022}).
\newblock


\bibitem[Trivedi et~al\mbox{.}(2022b)]%
        {trivedi2022musique}
\bibfield{author}{\bibinfo{person}{Harsh Trivedi}, \bibinfo{person}{Niranjan Balasubramanian}, \bibinfo{person}{Tushar Khot}, {and} \bibinfo{person}{Ashish Sabharwal}.} \bibinfo{year}{2022}\natexlab{b}.
\newblock \showarticletitle{MuSiQue: Multihop Questions via Single-hop Question Composition}.
\newblock \bibinfo{journal}{\emph{Transactions of the Association for Computational Linguistics}}  \bibinfo{volume}{10} (\bibinfo{year}{2022}), \bibinfo{pages}{539--554}.
\newblock


\bibitem[Wang et~al\mbox{.}(2025)]%
        {wang2025archragattributedcommunitybasedhierarchical}
\bibfield{author}{\bibinfo{person}{Shu Wang}, \bibinfo{person}{Yixiang Fang}, \bibinfo{person}{Yingli Zhou}, \bibinfo{person}{Xilin Liu}, {and} \bibinfo{person}{Yuchi Ma}.} \bibinfo{year}{2025}\natexlab{}.
\newblock \showarticletitle{ArchRAG: Attributed Community-based Hierarchical Retrieval-Augmented Generation}.
\newblock
\showeprint[arxiv]{2502.09891}~[cs.IR]
\urldef\tempurl%
\url{https://arxiv.org/abs/2502.09891}
\showURL{%
\tempurl}


\bibitem[Wang et~al\mbox{.}(2024)]%
        {wang2024large}
\bibfield{author}{\bibinfo{person}{Shen Wang}, \bibinfo{person}{Tianlong Xu}, \bibinfo{person}{Hang Li}, \bibinfo{person}{Chaoli Zhang}, \bibinfo{person}{Joleen Liang}, \bibinfo{person}{Jiliang Tang}, \bibinfo{person}{Philip~S Yu}, {and} \bibinfo{person}{Qingsong Wen}.} \bibinfo{year}{2024}\natexlab{}.
\newblock \showarticletitle{Large language models for education: A survey and outlook}.
\newblock \bibinfo{journal}{\emph{arXiv preprint arXiv:2403.18105}} (\bibinfo{year}{2024}).
\newblock


\bibitem[Wu et~al\mbox{.}(2024b)]%
        {wu2024medical}
\bibfield{author}{\bibinfo{person}{Junde Wu}, \bibinfo{person}{Jiayuan Zhu}, \bibinfo{person}{Yunli Qi}, \bibinfo{person}{Jingkun Chen}, \bibinfo{person}{Min Xu}, \bibinfo{person}{Filippo Menolascina}, {and} \bibinfo{person}{Vicente Grau}.} \bibinfo{year}{2024}\natexlab{b}.
\newblock \showarticletitle{Medical graph rag: Towards safe medical large language model via graph retrieval-augmented generation}.
\newblock \bibinfo{journal}{\emph{arXiv preprint arXiv:2408.04187}} (\bibinfo{year}{2024}).
\newblock


\bibitem[Wu et~al\mbox{.}(2024a)]%
        {wu2024retrieval}
\bibfield{author}{\bibinfo{person}{Shangyu Wu}, \bibinfo{person}{Ying Xiong}, \bibinfo{person}{Yufei Cui}, \bibinfo{person}{Haolun Wu}, \bibinfo{person}{Can Chen}, \bibinfo{person}{Ye Yuan}, \bibinfo{person}{Lianming Huang}, \bibinfo{person}{Xue Liu}, \bibinfo{person}{Tei-Wei Kuo}, \bibinfo{person}{Nan Guan}, {et~al\mbox{.}}} \bibinfo{year}{2024}\natexlab{a}.
\newblock \showarticletitle{Retrieval-augmented generation for natural language processing: A survey}.
\newblock \bibinfo{journal}{\emph{arXiv preprint arXiv:2407.13193}} (\bibinfo{year}{2024}).
\newblock


\bibitem[Xiang et~al\mbox{.}(2025)]%
        {xiang2025use}
\bibfield{author}{\bibinfo{person}{Zhishang Xiang}, \bibinfo{person}{Chuanjie Wu}, \bibinfo{person}{Qinggang Zhang}, \bibinfo{person}{Shengyuan Chen}, \bibinfo{person}{Zijin Hong}, \bibinfo{person}{Xiao Huang}, {and} \bibinfo{person}{Jinsong Su}.} \bibinfo{year}{2025}\natexlab{}.
\newblock \showarticletitle{When to use Graphs in RAG: A Comprehensive Analysis for Graph Retrieval-Augmented Generation}.
\newblock \bibinfo{journal}{\emph{arXiv preprint arXiv:2506.05690}} (\bibinfo{year}{2025}).
\newblock


\bibitem[Yang et~al\mbox{.}(2025)]%
        {yang2025qwen3}
\bibfield{author}{\bibinfo{person}{An Yang}, \bibinfo{person}{Anfeng Li}, \bibinfo{person}{Baosong Yang}, \bibinfo{person}{Beichen Zhang}, \bibinfo{person}{Binyuan Hui}, \bibinfo{person}{Bo Zheng}, \bibinfo{person}{Bowen Yu}, \bibinfo{person}{Chang Gao}, \bibinfo{person}{Chengen Huang}, \bibinfo{person}{Chenxu Lv}, {et~al\mbox{.}}} \bibinfo{year}{2025}\natexlab{}.
\newblock \showarticletitle{Qwen3 technical report}.
\newblock \bibinfo{journal}{\emph{arXiv preprint arXiv:2505.09388}} (\bibinfo{year}{2025}).
\newblock


\bibitem[Yang et~al\mbox{.}(2018)]%
        {yang2018HotpotQA}
\bibfield{author}{\bibinfo{person}{Zhilin Yang}, \bibinfo{person}{Peng Qi}, \bibinfo{person}{Saizheng Zhang}, \bibinfo{person}{Yoshua Bengio}, \bibinfo{person}{William~W Cohen}, \bibinfo{person}{Ruslan Salakhutdinov}, {and} \bibinfo{person}{Christopher~D Manning}.} \bibinfo{year}{2018}\natexlab{}.
\newblock \showarticletitle{HotpotQA: A dataset for diverse, explainable multi-hop question answering}.
\newblock \bibinfo{journal}{\emph{arXiv preprint arXiv:1809.09600}} (\bibinfo{year}{2018}).
\newblock


\bibitem[Yu et~al\mbox{.}(2024)]%
        {yu2024evaluation}
\bibfield{author}{\bibinfo{person}{Hao Yu}, \bibinfo{person}{Aoran Gan}, \bibinfo{person}{Kai Zhang}, \bibinfo{person}{Shiwei Tong}, \bibinfo{person}{Qi Liu}, {and} \bibinfo{person}{Zhaofeng Liu}.} \bibinfo{year}{2024}\natexlab{}.
\newblock \showarticletitle{Evaluation of retrieval-augmented generation: A survey}. In \bibinfo{booktitle}{\emph{CCF Conference on Big Data}}. Springer, \bibinfo{pages}{102--120}.
\newblock


\bibitem[Zhang et~al\mbox{.}(2025b)]%
        {zhang2025eraragefficientincrementalretrieval}
\bibfield{author}{\bibinfo{person}{Fangyuan Zhang}, \bibinfo{person}{Zhengjun Huang}, \bibinfo{person}{Yingli Zhou}, \bibinfo{person}{Qintian Guo}, \bibinfo{person}{Zhixun Li}, \bibinfo{person}{Wensheng Luo}, \bibinfo{person}{Di Jiang}, \bibinfo{person}{Yixiang Fang}, {and} \bibinfo{person}{Xiaofang Zhou}.} \bibinfo{year}{2025}\natexlab{b}.
\newblock \bibinfo{title}{EraRAG: Efficient and Incremental Retrieval Augmented Generation for Growing Corpora}.
\newblock
\showeprint[arxiv]{2506.20963}~[cs.IR]
\urldef\tempurl%
\url{https://arxiv.org/abs/2506.20963}
\showURL{%
\tempurl}


\bibitem[Zhang et~al\mbox{.}(2024a)]%
        {zhang2024sirerag}
\bibfield{author}{\bibinfo{person}{Nan Zhang}, \bibinfo{person}{Prafulla~Kumar Choubey}, \bibinfo{person}{Alexander Fabbri}, \bibinfo{person}{Gabriel Bernadett-Shapiro}, \bibinfo{person}{Rui Zhang}, \bibinfo{person}{Prasenjit Mitra}, \bibinfo{person}{Caiming Xiong}, {and} \bibinfo{person}{Chien-Sheng Wu}.} \bibinfo{year}{2024}\natexlab{a}.
\newblock \showarticletitle{SiReRAG: Indexing Similar and Related Information for Multihop Reasoning}.
\newblock \bibinfo{journal}{\emph{arXiv preprint arXiv:2412.06206}} (\bibinfo{year}{2024}).
\newblock


\bibitem[Zhang et~al\mbox{.}(2025a)]%
        {zhang2025survey}
\bibfield{author}{\bibinfo{person}{Qinggang Zhang}, \bibinfo{person}{Shengyuan Chen}, \bibinfo{person}{Yuanchen Bei}, \bibinfo{person}{Zheng Yuan}, \bibinfo{person}{Huachi Zhou}, \bibinfo{person}{Zijin Hong}, \bibinfo{person}{Junnan Dong}, \bibinfo{person}{Hao Chen}, \bibinfo{person}{Yi Chang}, {and} \bibinfo{person}{Xiao Huang}.} \bibinfo{year}{2025}\natexlab{a}.
\newblock \showarticletitle{A Survey of Graph Retrieval-Augmented Generation for Customized Large Language Models}.
\newblock \bibinfo{journal}{\emph{arXiv preprint arXiv:2501.13958}} (\bibinfo{year}{2025}).
\newblock


\bibitem[Zhang et~al\mbox{.}(2024b)]%
        {zhang2024there}
\bibfield{author}{\bibinfo{person}{Yunan Zhang}, \bibinfo{person}{Shige Liu}, {and} \bibinfo{person}{Jianguo Wang}.} \bibinfo{year}{2024}\natexlab{b}.
\newblock \showarticletitle{Are there fundamental limitations in supporting vector data management in relational databases? A case study of PostgreSQL}.
\newblock \bibinfo{journal}{\emph{2024 IEEE 40th International Conference on Data Engineering (ICDE)}} (\bibinfo{year}{2024}), \bibinfo{pages}{3640--3653}.
\newblock


\bibitem[Zhao et~al\mbox{.}(2024)]%
        {zhao2024retrieval}
\bibfield{author}{\bibinfo{person}{Penghao Zhao}, \bibinfo{person}{Hailin Zhang}, \bibinfo{person}{Qinhan Yu}, \bibinfo{person}{Zhengren Wang}, \bibinfo{person}{Yunteng Geng}, \bibinfo{person}{Fangcheng Fu}, \bibinfo{person}{Ling Yang}, \bibinfo{person}{Wentao Zhang}, \bibinfo{person}{Jie Jiang}, {and} \bibinfo{person}{Bin Cui}.} \bibinfo{year}{2024}\natexlab{}.
\newblock \showarticletitle{Retrieval-augmented generation for ai-generated content: A survey}.
\newblock \bibinfo{journal}{\emph{arXiv preprint arXiv:2402.19473}} (\bibinfo{year}{2024}).
\newblock


\bibitem[Zhou et~al\mbox{.}(2025)]%
        {zhou2025depth}
\bibfield{author}{\bibinfo{person}{Yingli Zhou}, \bibinfo{person}{Yaodong Su}, \bibinfo{person}{Youran Sun}, \bibinfo{person}{Shu Wang}, \bibinfo{person}{Taotao Wang}, \bibinfo{person}{Runyuan He}, \bibinfo{person}{Yongwei Zhang}, \bibinfo{person}{Sicong Liang}, \bibinfo{person}{Xilin Liu}, \bibinfo{person}{Yuchi Ma}, {et~al\mbox{.}}} \bibinfo{year}{2025}\natexlab{}.
\newblock \showarticletitle{In-depth Analysis of Graph-based RAG in a Unified Framework}.
\newblock \bibinfo{journal}{\emph{arXiv preprint arXiv:2503.04338}} (\bibinfo{year}{2025}).
\newblock


\end{thebibliography}
  
\appendix

\section{Supplementary Details}
\label{appendix}
\subsection{Experimental Settings}

\label{Experimental_Settings}
For token cost comparison between offline indexing and online retrieval using LLaMA3.0-8B shown in Figure \ref{fig:token_compare_optimized} and \ref{fig:query_token}, we account for both prompt and completion tokens. For methods requiring LLM-based query preprocessing (e.g., keyword or entity extraction), these additional token costs are included in our calculations. During the LLM generation phase, we employ a strategy by duplicating the input query to the end of the original prompt. This design explicitly mitigates the "lost in the middle" effect \cite{liu2023lost} observed in approaches like LightRAG, where placing the question before lengthy context can cause the model to overlook the task objective \cite{ram2023context, liu2023lost}. Furthermore, inspired by the prompt designs of SIRERAG \cite{zhang2024sirerag} and KETRAG \cite{huang2025ket}, which include phrases like "Answer this question as short as possible" to prevent verbose or irrelevant outputs, we uniformly add this instruction to the final prompt of each method. This ensures concise responses, maintaining evaluation fairness and highlighting our method's effectiveness. For other hyperparameters of each method, we follow the original settings in their available codes. For NLP pre-processing, we employ the following tools: cl100k\_base from tiktoken library \cite{openai_tik} for word tokenization and counting, NLTK's sent\_tokenize \cite{Bird_Natural_Language_Processing_2009} for sentence segmentation, and spaCy \cite{Honnibal_spaCy_Industrial-strength_Natural_2020} for named entity recognition. For LLM inference, we use LLaMA3.0-8B \cite{grattafiori2024llama} and Qwen2.5-32B \cite{bai2023qwen}, with a maximum token limit of 8,000. In top-$K$ selection tasks, we set $K = 5$ to accommodate token constraints. The embedding model for vector search is BGE-M3 \cite{chen2024bge}, while re-ranking is performed using the lightweight yet powerful BGE-Reranker-v2-M3 \cite{chen2024bge}. As for our proposed approach, Clue-RAG employs the following default configurations: the number of most relevant results $K$ is set to 3, search depth $D$ to 3, beam size $M$ per depth to 5, and returned candidate chunks $N$ to 5, ensuring that all methods ultimately retrieve the same number of query-relevant chunks (i.e., 5) for fair comparison. By default, we set the value of $\alpha$ to 1 if it is not explicitly specified. Additionally, BLEU score \cite{papineni2002bleu} serves as the core selection metric for relevance scoring.

\subsection{Dataset Details}

\label{algorithm_appendix}
Table \ref{tab:dataset} presents the key statistics for the three benchmark datasets used in our study: MuSiQue, HotpotQA, and 2Wiki, including corpora size, number of questions, and total token count. Table \ref{tab:count_rag_1.0} reports the node counts for the Clue-RAG-1.0 graph index, detailing the quantities of text nodes, knowledge unit nodes, and entity nodes for each dataset. Figure \ref{fig:hyperparameter_appendix} presents the performance under different hyperparameter settings on HotpotQA and 2Wiki datasets.


\begin{table}[H]
\centering
\caption{Details of three benchmark datasets.}
\begin{tabular}{lccc}
\hline
\textbf{Dataset} & \textbf{MuSiQue} & \textbf{HotpotQA} & \textbf{2Wiki} \\
\hline
{\tt Questions} & 1,000 & 1,000 & 1,000 \\
{\tt Passages} & 11,656 & 9,221 & 6,119 \\
{\tt Tokens} & 1,281,422 & 1,239,838 & 640,205 \\
\hline

\end{tabular}

\label{tab:dataset}
\end{table}

\begin{table}[H]
\centering
\caption{Statistics of Clue-RAG-1.0's graph index in LLaMA3.0-8B and Qwen2.5-32B.}
\begin{tabular}{llccc}
\hline
\textbf{Model} & \textbf{Nodes} & \textbf{MuSiQue} & \textbf{HotpotQA} & \textbf{2Wiki} \\
\hline
\multirow{3}{*}{\makecell[c]{LLaMA \\ 3.0-8B} }
& Texts &  11,656 & 9,221 & 6,119 \\
& Knowledge & 108,513 & 100,789 & 52,149 \\
& Entities & 222,940 & 209,869 & 123,116 \\

\hline


\multirow{3}{*}{\makecell[c]{Qwen \\ 2.5-32B} }
& Texts &  11,656 & 9,221 & 6,119 \\
& Knowledge & 114,256 & 108,957 & 57,442 \\
& Entities & 238,954 & 229,762 & 136,966 \\

\hline
\end{tabular}
\label{tab:count_rag_1.0}
\end{table}

\definecolor{ch1}{HTML}{c7522a} 
\definecolor{ch2}{HTML}{008585} 
\begin{figure}[htbp] 
    \centering
\begin{tikzpicture}
    \begin{groupplot}[
        group style={
            group size=4 by 2,
            horizontal sep=0.5cm,
        },
        width=0.18\textwidth,
        height=0.15\textwidth,
        tick label style={font=\scriptsize},   
    ]

    \nextgroupplot[title={}, ymin=63, ymax=65, ytick={63, 64, 65},ylabel near ticks,
    xmin=3, xmax=9,
    xtick={3,5,7,9},xlabel={},]
    
    \addplot[
            ch1, 
            mark=*, 
            mark options={solid, fill=white},  
            line width=1.0pt,                 
        ] coordinates {(3,64.02)(5,63.23)(7,63.83)(9,63.31)};
        
    \addplot[
            ch2, 
            mark=triangle*, 
            mark options={solid, fill=white},  
            line width=1.0pt,                 
        ] coordinates {(3,64.70)(5,64.40)(7,65.00)(9,64.60)};

    \nextgroupplot[ymin=60, ymax=68, ytick={60, 64, 68},ylabel near ticks,
    xmin=3, xmax=9,
    xtick={3,5,7,9},xlabel={}]

    \addplot[
            ch1, 
            mark=*, 
            mark options={solid, fill=white},  
            line width=1.0pt,                 
        ] coordinates {(3,60.28)(5,64.02)(7,64.86)(9,64.57)};
        
    \addplot[
            ch2, 
            mark=triangle*, 
            mark options={solid, fill=white},  
            line width=1.0pt,                 
        ] coordinates {(3,61.10)(5,64.70)(7,66.20)(9,66.20)};

\nextgroupplot[title={HotpotQA},ymin=55, ymax=65, ytick={55, 60, 65},ylabel near ticks,
    xmin=1, xmax=4,
    xtick={1,2,3,4},xlabel={},title style={yshift=-1mm,xshift=-10mm,}]

\addplot[
    ch1, 
    mark=*, 
    mark options={solid, fill=white},  
    line width=1.0pt                  
] coordinates {
    (1,55.87)
    (2,62.02)
    (3,64.02)
    (4,63.44)
};


\addplot[
    ch2,
    mark=triangle*,
    mark options={solid, fill=white},  
    line width=1.0pt                  
] coordinates {
    (1,56.50)
    (2,62.90)
    (3,64.70)
    (4,64.10)
};


\nextgroupplot[ymin=62, ymax=66, ytick={62, 64, 66},ylabel near ticks,
    xmin=0.0, xmax=1.0,
    xtick={0.0, 0.25, 0.5, 0.75, 1.0},xlabel={}]

\addplot[
    ch1, 
    mark=*, 
    mark options={solid, fill=white},  
    line width=1.0pt                  
] coordinates {
    (0.0,62.14)
    (0.25,62.47)
    (0.5,63.44)
    (0.75,64.52)
    (1.0,64.02)
};

\addplot[
    ch2,
    mark=triangle*,
    mark options={solid, fill=white},  
    line width=1.0pt                  
] coordinates {
    (0.0,63.50)
    (0.25,63.20)
    (0.5,64.10)
    (0.75,65.40)
    (1.0,64.70)
};

\nextgroupplot[ymin=52, ymax=62, ytick={52, 57, 62},ylabel near ticks,title style={yshift=-5pt},xmin=3, xmax=9,
    xtick={3,5,7,9}, xlabel={\textbf{(a) Top-$\mathbf{\textit{K}}$}}]
    \addplot[
            ch1, 
            mark=*, 
            mark options={solid, fill=white},  
            line width=1.0pt,                 
        ] coordinates {(3,55.20)(5,53.63)(7,53.02)(9,52.63)};
        
    \addplot[
            ch2, 
            mark=triangle*, 
            mark options={solid, fill=white},  
            line width=1.0pt,                 
        ] coordinates {(3,61.00)(5,59.30)(7,59.00)(9,58.00)};

\nextgroupplot[ymin=50, ymax=62, ytick={50, 56, 62},ylabel near ticks, xmin=3, xmax=9,
    xtick={3,5,7,9},xlabel={\textbf{(b) $\mathbf{\textit{M}}$}}]

    \addplot[
            ch1, 
            mark=*, 
            mark options={solid, fill=white},  
            line width=1.0pt,                 
        ] coordinates {(3,50.36)(5,55.20)(7,55.81)(9,54.92)};
        
    \addplot[
            ch2, 
            mark=triangle*, 
            mark options={solid, fill=white},  
            line width=1.0pt,                 
        ] coordinates {(3,56.20)(5,61.00)(7,62)(9,61.50)};

\nextgroupplot[title={2Wiki},ymin=50, ymax=62, ytick={50, 56, 62}, ylabel near ticks, xmin=1, xmax=4,xtick={1,2,3,4}, title style={yshift=-1mm,xshift=-10mm,},xlabel={\textbf{(c) $\mathbf{\textit{D}}$}}]

\addplot[
    ch1, 
    mark=*, 
    mark options={solid, fill=white},  
    line width=1.0pt                  
] coordinates {
    (1,50.59)
    (2,54.51)
    (3,55.20)
    (4,53.79)
};

\addplot[
    ch2,
    mark=triangle*,
    mark options={solid, fill=white},  
    line width=1.0pt                  
] coordinates {
    (1,55.10)
    (2,60.30)
    (3,61.00)
    (4,59.20)
};

\nextgroupplot[ymin=49, ymax=61, ytick={49, 55, 61}, ylabel near ticks,xmin=0.0, xmax=1.0,
    xtick={0.0, 0.25, 0.5, 0.75, 1.0},xlabel={\textbf{(d) $\mathbf{\alpha}$}}]

\addplot[
    ch1,
    mark=*,
    mark options={solid, fill=white},  
    line width=1.0pt                  
] coordinates {
    (0.0,49.47)
    (0.25,51.23)
    (0.5,52.49)
    (0.75,53.62)
    (1.0,55.20)
};

\addplot[
    ch2,
    mark=triangle*,
    mark options={solid, fill=white},  
    line width=1.0pt                  
] coordinates {
    (0.0,54.70)
    (0.25,57.40)
    (0.5,58.80)
    (0.75,59.40)
    (1.0,61.00)
};

    \end{groupplot}
\end{tikzpicture}
\caption{Performance analysis under different hyperparameter settings on HotpotQA and 2Wiki datasets.}

\label{fig:hyperparameter_appendix}
\end{figure}
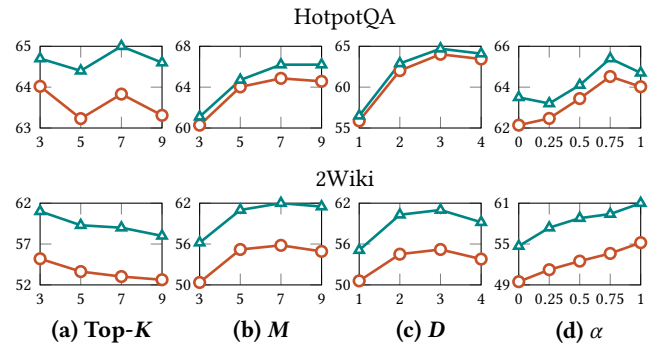

\subsection{Prompt Templates}     
Figure \ref{fig:propositions} and \ref{fig:generation} illustrate the LLM instruction prompts used for extracting knowledge units and answer generations, respectively.

\begin{figure*}[] 
\begin{AIbox}{Prompt for generating knowledge units}
{\bf Prompt:} \\
{
Decompose the content into clear knowledge units, ensuring they are interpretable independently of their original context:
\begin{enumerate}[leftmargin=*]
    \item Sentence Simplification: Break compound sentences into simpler, individual sentences. Whenever possible, retain the original phrasing from the input text.
    \item Entity and Description Separation: For named entities that are accompanied by descriptive information, separate the descriptive details into a distinct knowledge unit. Ensure each knowledge unit represents a single, clear fact.
    \item Pronoun Resolution: Replace all pronouns (e.g., "it", "they", "this") with explicit references, using full taxonomic names or clear identifiers. Always use "[entity]'s [property]" construction.
    \item Output Format: Present the resulting knowledge units as a list of strings, formatted in JSON.
\end{enumerate}

{\bf EXAMPLE-1:}
\newline 
Input: 
\newline
Jesúfas Aranguren. His 13-year professional career was solely associated with Athletic Bilbao, with which he played in nearly 400 official games, winning two Copa del Rey trophies.

\begin{verbatim}
Output:
{
    "knowledge units": [
        "Jesús Aranguren had a 13-year professional career.",
        "Jesús Aranguren's professional career was solely associated with Athletic Bilbao.",
        "Athletic Bilbao is a football club.",
        "Jesús Aranguren played for Athletic Bilbao in nearly 400 official games.",
        "Jesús Aranguren won two Copa del Rey trophies with Athletic Bilbao.",
    ]
}
\end{verbatim}

{\bf EXAMPLE-2:}
\newline 
Input:
\newline
Ophrys apifera. Ophrys apifera grows to a height of 15 -- 50 centimetres (6 -- 20 in). This hardy orchid develops small rosettes of leaves in autumn. They continue to grow slowly during winter. Basal leaves are ovate or oblong - lanceolate, upper leaves and bracts are ovate - lanceolate and sheathing. The plant blooms from mid-April to July producing a spike composed from one to twelve flowers. The flowers have large sepals, with a central green rib and their colour varies from white to pink, while petals are short, pubescent, yellow to greenish. The labellum is trilobed, with two pronounced humps on the hairy lateral lobes, the median lobe is hairy and similar to the abdomen of a bee. It is quite variable in the pattern of coloration, but usually brownish - red with yellow markings. The gynostegium is at right angles, with an elongated apex.

\begin{verbatim}
Output: 
{ 
    "knowledge units": [
        "Ophrys apifera grows to a height of 15-50 centimetres (6-20 in)", 
        "Ophrys apifera is a hardy orchid", 
        "Ophrys apifera develops small rosettes of leaves in autumn", 
        "The leaves of Ophrys apifera continue to grow slowly during winter", 
        "The basal leaves of Ophrys apifera are ovate or oblong-lanceolate", 
        "The upper leaves and bracts of Ophrys apifera are ovate-lanceolate and sheathing", 
        "Ophrys apifera blooms from mid-April to July", 
        "Ophrys apifera produces a spike composed of one to twelve flowers", 
        "The flowers of Ophrys apifera have large sepals with a central green rib", 
        "The flowers of Ophrys apifera vary in colour from white to pink", 
        "The petals of Ophrys apifera are short, pubescent, and yellow to greenish", 
        "The labellum of Ophrys apifera is trilobed with two pronounced humps on the hairy lateral lobes", 
        "The median lobe of Ophrys apifera's labellum is hairy and resembles a bee's abdomen", 
        "The coloration pattern of Ophrys apifera is variable but usually brownish-red with yellow markings", 
        "The gynostegium of Ophrys apifera is at right angles with an elongated apex" ,
    ]
}

\end{verbatim}

JUST OUTPUT THE RESULTS IN JSON FORMAT!
\newline
Input: \{passage\}
 \newline
 Output:

}

\end{AIbox} 
\caption{The prompt for generating knowledge units.}
\label{fig:propositions}
\end{figure*}
\begin{figure*}[] 
\begin{AIbox}{Prompt for answer generation}
{\bf Prompt:} \\
{
Your goal is to give the best full answer to question the user input according to the given context below.

Given Context: \{context\}

Give the best full answer to question :\{question\}

Answer this question in as fewer number of words as possible.
}

\end{AIbox} 
\caption{The prompt for generating answer.}
\label{fig:generation}
\end{figure*}

\end{document}